\documentclass[twocolumn,aps,prx,superscriptaddress]{revtex4-1}
\usepackage{setspace}
\usepackage{graphicx}
\usepackage{subfigure}
\usepackage{amsmath,amssymb}
\usepackage{color}
\usepackage{enumitem}
\usepackage{soul}
\usepackage{xr-hyper}
\usepackage{hyperref}
\newcommand{\be}{\begin{equation}}
\newcommand{\ee}{\end{equation}}

\begin{document}
\begin{singlespace}

% The title, author and date are defined here, for later formatting with
% the \maketitle command
\title{Phase transitions in \emph{in vivo} or \emph{in vitro} populations of spiking neurons belong to different universality classes}
%\title{Circuit environment, connectivity statistics, and feedback inhibition control the universality classes of phase transitions in spiking neural populations}
%\title{Circuit environment and feedback inhibition control the universality classes of phase transitions in spiking neural populations}

\author{Braden A. W. Brinkman}
\affiliation{Department of Neurobiology and Behavior, Stony Brook University, Stony Brook, NY, 11794, USA}
\date{\today}

\begin{abstract}
The ``critical brain hypothesis'' posits that neural circuitry may be tuned close to a ``critical point'' or ``phase transition''---a boundary between different operating regimes of the circuit. 
The renormalization group and theory of critical phenomena explain how systems tuned to a critical point display scale invariance due to fluctuations in activity spanning a wide range of time or spatial scales. 
In the brain this scale invariance has been hypothesized to have several computational benefits, including increased collective sensitivity to changes in input and robust propagation of information across a circuit. 
However, our theoretical understanding of critical phenomena in neural circuitry is limited because standard renormalization group methods apply to systems with either highly organized or completely random connections. 
Connections between neurons lie between these extremes, and may be either excitatory (positive) or inhibitory (negative), but not both. 
In this work we develop a renormalization group method that applies to models of spiking neural populations with some realistic biological constraints on connectivity, and derive a scaling theory for the statistics of neural activity when the population is tuned to a critical point. 
We show that the scaling theories differ for models of \emph{in vitro} versus \emph{in vivo} circuits---they belong to different ``universality classes''---and that both may exhibit ``anomalous'' scaling at a critical balance of inhibition and excitation.
We verify our theoretical results on simulations of neural activity data, and discuss how our scaling theory can be further extended and applied to real neural data.

\end{abstract}
\maketitle

%%%%%%%%%%%%%%%%%%%%%%%%%%%%%%%%
% Intro
%%%%%%%%%%%%%%%%%%%%%%%%%%%%%%%%

There is little hope of understanding how each of the $\mathcal O(10^{11})$ neurons contributes to the functions of the brain \cite{KassAnnRevStats2018}. 
Even individual brain regions contain millions of neurons \cite{wandell1995foundations,collins2010neuron}, more than can be individually mapped out, but enough that the tools of statistical physics can be applied to understand how collective patterns of neural activity may contribute to brain function.
Indeed, experimental work has demonstrated that neural circuitry can operate in many different regimes of collective activity \cite{rabinovich2006dynamical,wimmer2014bump,kim2017ring,beggs2003neuronal,FriedmanPRL2012}. 
Theoretical and computational analyses of these collective dynamics suggest that transitions between different operating regimes may be sharp, akin to phase transitions observed in statistical physics \cite{rabinovich2006dynamical,zhang1996representation,laing2001stationary,kim2017ring,ermentrout1979mathematical,bressloff2010metastable,butler2012evolutionary,SompolinskyPRL1988,dahmen2019second,buice2007field,kadmon2015transition}. 
The theory of critical phenomena predicts that at a phase transition the statistical fluctuations of a system span many orders of magnitude in space and time, and the system displays approximate scale invariance \cite{GoldenfeldBook1992}.
In the brain, scale invariance would lead not only to power-law scaling in neural activity, but \emph{scaling collapse}, in which activity data collected under different conditions can be systematically rescaled to fall onto a single universal curve.
Such signatures of criticality have been observed in neural data from the retina \cite{tkavcik2015thermodynamics}, visual cortex \cite{ma2019cortical,xu2024sleep}, hippocampus \cite{meshulam2019coarse}, and other cortical areas \cite{cocchi2017criticality}.
These observations have lead neuroscientists to hypothesize that circuitry in the brain is actively maintained close to critical points---the dividing lines between phases; this has become known as the ``critical brain hypothesis'' \cite{shew2011information,BeggsFrontPhysio2012,FriedmanPRL2012,ShewTheNeuro2013}. 
Proponents of this hypothesis argue that these scale-spanning fluctuations could benefit brain function by minimizing circuit reaction times to perturbations, facilitating switches between computations and maximizing information transferred \cite{BeggsFrontPhysio2012}.

However, our understanding of critical phenomena and scaling in neural systems is largely phenomenological, based on analogies with well-studied systems from physics that lack many of the biological features of neural circuits, such as complex network structure and distinct excitatory and inhibitory cell types.
This makes it difficult to resolve apparently inconsistent measurements of signatures of criticality across different brain areas.
For example, some analyses of neural data appear to suggest that power law exponents are of the ``mean-field'' type, predictable by standard dimensional analysis, while other studies point towards anomalous exponents that that deviate from the mean-field predictions \cite{FriedmanPRL2012,yaghoubi2018neuronal,fontenele2019criticality}.
Anomalous power law exponents come in sets corresponding to different ``universality classes,'' where the universality class of a system is characterized by symmetries of the dynamics and statistical distributions, and, in lattices and continuous media, the dimension of the system.
While universality classes and anomalous scaling are well understood theoretically in lattices and continuous media, the situation on more general networks remains an open problem.

The modern understanding of critical phenomena, and the origin of anomalous scaling, is based on the renormalization group (RG).
The RG is a framework for organizing activity into a hierarchy of scales and determining how statistical fluctuations at each scale contribute to the overall statistics of a system.
In soft condensed matter physics, these scales are typically distance and time, and the RG reveals how microscopic details influence dynamics and statistics on long spatial and temporal scales.
However, it is not clear what the appropriate scales are in neural circuits, as spatial distance between neurons does not necessarily reflect the influence they have on each other through chains of synaptic connections. 

Recent work has used ideas from the RG to devise phenomenological schemes for analyzing data, determining that a system is critical if the data can be shown to be approximately scale invariant under repeated coarse-graining of the principal components of neural covariances \cite{bradde2017pca,meshulam2019coarse,nicoletti2020scaling,ponce2023critical} or in time \cite{sooter2024cortex}.
However, a theoretical understanding of neural systems through the lens of the RG has so far been restricted to models from statistical physics that are re-interpreted in terms of of coarse-grained neural signals, such as active/inactive units \cite{buice2007field,bradde2017pca}, or in networks of neurons described by firing rates \cite{stapmanns2020self,tiberi2022gell}, rather than populations of neurons that emit spikes, the fundamental unit of communication in neural circuits.

In this work we establish this missing theoretical foundation and develop a scaling theory for the relaxation of neural activity to a steady state.
To the author's knowledge, the results obtained follow from the first full theoretic RG analysis of neural populations with leaky integrate-and-fire spiking dynamics.
We consider models of both \emph{in vitro} circuits---slices of tissue removed from the brain---and \emph{in vivo} circuits---recordings directly from in-tact brain tissues---and show that they belong to the directed percolation and Ising model universality classes, respectively.
We first perform a mean-field analysis of the model to show that it predicts two different types of phase transitions, for which we derive the scaling collapse relations and illustrate the idea of data collapse  (Sec.~\ref{sec:spikingmodel}).
We then give and overview of how fluctuations change the mean-field picture and the scaling relations. 
We show that simulated data of several networks can indeed be collapsed onto universal scaling forms, some yielding mean-field exponents and others anomalous scaling (Sec.~\ref{sec:applications}). 
We end this report by discussing the implications of this work for current theoretical and experimental investigations of collective activity in spiking networks, both near and away from phase transitions (Sec.~\ref{sec:discussion}).

\begin{figure*}
 \centering
\includegraphics[scale=0.16]{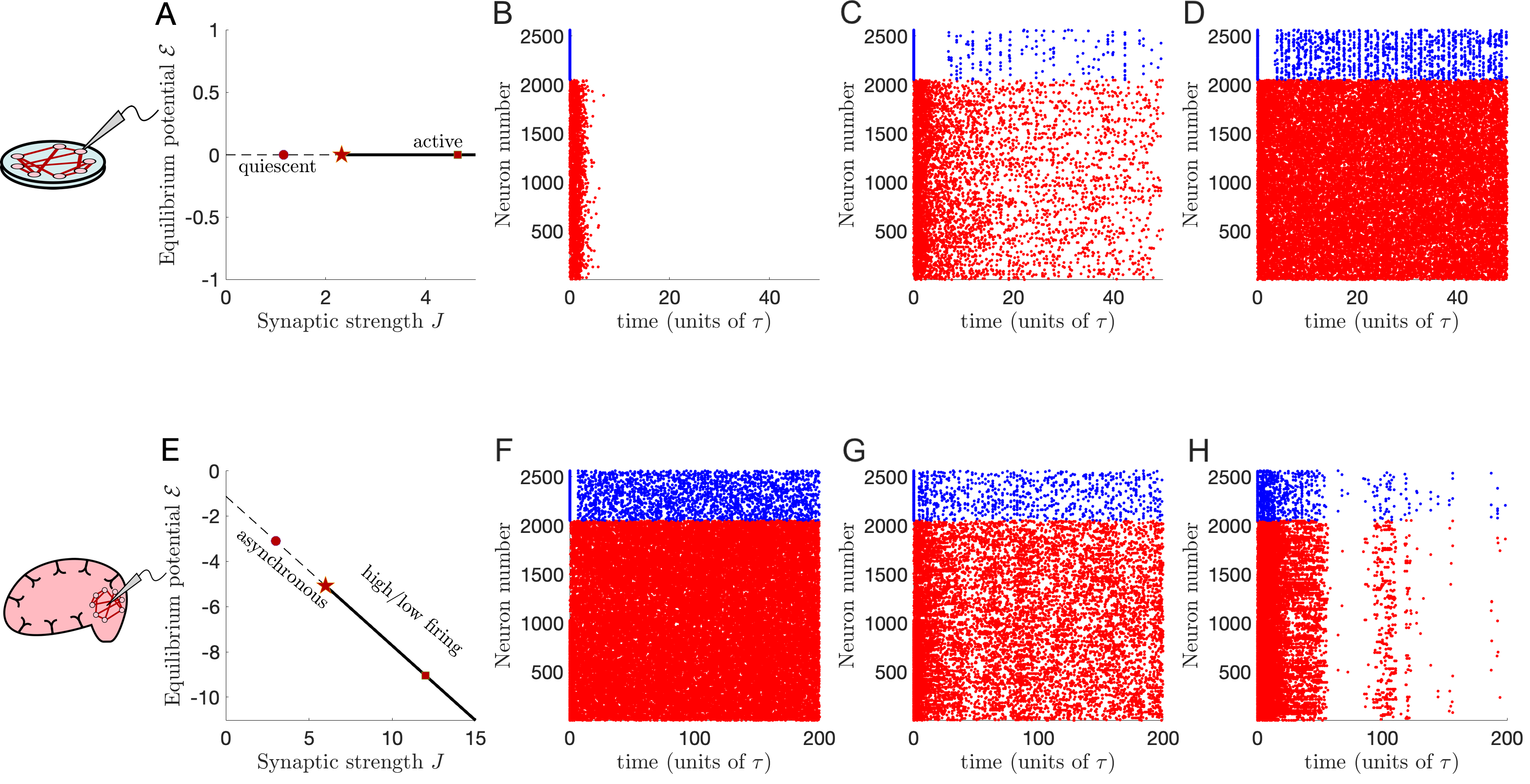}
  \caption{\textbf{Phase transitions in \emph{in vitro} versus \emph{in vivo} neural populations.} Neural activity may differ between recorded from tissue maintained or grown in a pitri dish (``\emph{in vitro}''; top row) or recorded directly from neurons in a living organism (``\emph{in vivo}''; bottom row). These differences partly reflect external input: \emph{in vitro} tissue may require experimenter-provided stimulation to maintain the activity of neurons, while \emph{in vivo} neurons are constantly bombarded with input from other brain areas or body systems, leading to spontaneous activity. In both cases qualitative changes in population-level activity may be observed as properties of the network, such as the overall strength of synaptic connections, are modulated. \textbf{A.} Phase diagram of an \emph{in vitro} network: if the equilibrium resting potential of the neurons is perturbed (e.g., due to external tonic current input), the network's firing can be suppressed ($\mathcal E < 0$) or promoted ($\mathcal E > 0$). At the equilibrium potential $\mathcal E = 0$ (normalized units) the network activity will decay away if the strength of synaptic connections is less than a critical value $J_c$. For synaptic strengths $J > J_c$ the network activity is self-sustaining. At the critical value $J_c$ the activity decays to quiescence, but very slowly. \textbf{B.} Example raster plot of spiking activity in a network with subcritical $J = J_c/2$ (circle), along the line $\mathcal E = 0$, showing a fast decay of activity. \textbf{C.} Spiking activity in a network at the approximate critical point $J = J_c$ (star), showing slow decay of activity. \textbf{D.} Spiking activity in a supercritical network with $J = 2J_c$ (square), showing sustained activity. \textbf{E.} Phase diagram of an \emph{in vivo} network: perturbing the equilibrium resting potential will increase or decrease neural firing. Along a critical line $\mathcal E = \mathcal E_c$ (dashed diagonal line) the network will fire asynchronously. For synaptic strengths $J > J_c$ there exist states of low or high firing, which the network can transition spontaneously between in finite networks. \textbf{F.} Example raster plot of spiking activity in a network with subcritical $J = J_c/2$ (circle), along the approximate critical line $\mathcal E = \mathcal E_c$, showing asynchronous activity. \textbf{G.} Spiking activity in a network at the approximate critical point $J = J_c$ (star), showing intermittent high and low spiking activity activity. \textbf{H.} Spiking activity in a supercritical network with $J = 2J_c$ (square), showing apparent transient metastable transitions between high and low firing rate states that are possible in finite-sized network simulations. Excitatory neurons are colored red, inhibitory neurons are colored blue.}
  \label{fig:introschematic}
% \vspace{-.1in}
\end{figure*}

%%%%%%%%%%%%%%%%%%%%%%%%%%%%%%%%
%% RESULTS 
%%%%%%%%%%%%%%%%%%%%%%%%%%%%%%%%

\section{Spiking network model}
\label{sec:spikingmodel}

We consider a network of $N$ neurons that stochastically fire action potentials, which we refer to as ``spikes.'' The probability that neuron $i$ fires $\dot{n}_i(t)dt$ spikes within a small window $[t, t+dt]$ is given by a counting process with expected value $\phi(V_i(t))dt$, where $\phi(V)$ is a non-negative firing rate nonlinearity, conditioned on the current value of the membrane potential $V_i(t)$. We assume $\phi(V)$ is the same for all neurons, and for definiteness we will take the counting process to be Poisson or Bernoulli, though the properties of the phase transitions should not depend on this specific choice. 

The membrane potential of each neuron obeys leaky dynamics,
\begin{align}
    \tau \frac{dV_i}{dt} &= -(V_i - \mathcal E) + \sum_{j=1}^N J_{ij} \dot{n}_j(t),     \label{eqn:LIF} \\
    \dot{n}_i(t)dt &\sim {\rm Poiss}[\phi(V_i(t))dt] \label{eqn:poiss}
\end{align}
where $\tau$ is the membrane time constant, $\mathcal E$ is the equilibrium potential of the neuron in the absence of input, and $J_{ij} \equiv J(A_{ij} - \delta_{ij})$ is the weight of the synaptic connection from pre-synaptic neuron $j$ to post-synaptic neuron $i$. 
The synaptic weights are characterized by a strength $J$, and the connections between neurons are encoded by the adjacency matrix $A_{ij}$, which is $1$ if neuron $j$ connects onto neuron $i$ and $0$ otherwise.
We take $J_{ii} = -J$ to be negative in order to implement a ``soft'' refractory effect, resetting a neuron's membrane potential by a fixed amount $-J$ after each spike.
For simplicity, we model the synaptic input as an instantaneous impulse, referred to as a ``pulse coupled'' network. 
We focus on symmetric networks $J_{ij} = J_{ji}$ with a largest real-valued eigenvalue $J\Lambda_{\rm max}$ associated with a homogeneous eigenmode, where $\Lambda_{\rm max}$ is the maximum eigenvalue of $A_{ij}-\delta_{ij}$. 
That is, phase transitions in these networks will correspond to pattern formation out of a homogeneous state of activity.
While real networks are not homogeneous and symmetric, reciprocal pairs of connections are more common than expected for random networks, and tend to be stronger than uni-directional connections \cite{song2005highly}. 
We interpret our networks as an approximation in which unidirectional connections and and variance in the synaptic weights can be neglected. 
Despite the restriction of a leading homogeneous mode, this encompasses a broad and important class of models and networks, such as  bump attractor models in neuroscience \cite{wimmer2014bump,kim2017ring} and, more generally, diffusion on networks \cite{villegas2023laplacian}.

In this work we consider two nonlinearities corresponding to two types of networks, \emph{in vitro} and \emph{in vivo} networks:
\begin{align}
\phi(V) &= \left\{ \begin{array}{c c} \lambda_0\left\lfloor \frac{1}{1+e^{-(V-\theta)/V_s}} - r_0\right\rfloor_+&,~{\rm \emph{in~vitro}~networks}\\  \frac{\lambda_0}{1+e^{-(V-\theta)/V_s}}&,~{\rm \emph{in~vivo}~networks}\end{array} \right. ,
\label{eqn:phis}
\end{align}
where $V_s$ sets the slope of the nonlinearities and $\lambda_0$ sets the maximum firing rate of the neurons (equal to $\lambda_0(1-r_0)$ for \emph{in vitro} networks and $\lambda_0$ for \emph{in vivo} networks). 
We will choose units such that $V_s = \lambda_0 = 1$. 
The soft threshold $\theta$ is the value that the membrane potential needs to exceed in order for a neuron to have an increased probability of firing a spike. 

The key difference between the \emph{in vitro} and \emph{in vivo} network nonlinearities is the presence of the rectification in the \emph{in vitro} network nonlinearity, which creates the absorbing state: when $V \leq \theta - \ln(r_0^{-1}-1)$, due to the shift down by $r_0$, neurons in an \emph{in vitro} network will not fire. 
If all neurons' membrane potentials are below this threshold the network will be completely quiescent and cannot fire spikes without further external input, so this quiescent phase constitutes an absorbing state of the network.
In contrast, the instantaneous firing rates of neurons in \emph{in vivo} networks never vanish for any finite input, so there is always some probability that a neuron can fire a spike, even if that probability is small.
Other important features of these networks are the saturation of the instantaneous firing rate and the concavity of the nonlinearity.
For example, if the nonlinearity is unbounded the network may become unstable for $J > J_c$, leading to runaway excitation of the neurons.

In this work we probe the behavior of the network by potentiating or suppressing the membrane potentials of the entire population, and then allowing the network to relax back to a steady state. 
If the network is tuned to a critical point this relaxation will follow a power law.
This procedure would approximate an experimental set-up in which a wide area of neural tissue is optogenetically stimulated or suppressed uniformly.
This procedure is similar to experimental setups investigating neural ``avalanches''---cascades of neural activity typically originating from a single neuron.
However, by potentiating or suppressing the entire network, there is no single neuron that triggers the cascade of activity, allowing us to avoid ambiguity in defining avalanches in the spontaneously active \emph{in vivo} networks, in which it can be unclear whether multiple clusters of activity are really independent events or just non-local parts of a single avalanche.

In Fig.~\ref{fig:introschematic} we display two types of phase transitions this stochastic spiking model can exhibit.
The first type involves a transition between a quiescent, inactive state, and a self-sustained active state (Fig.~\ref{fig:introschematic}A-D).
This is an appropriate model for \emph{in vitro} networks, tissue removed from the brain and maintained or cultured in a dish.
Such networks receive little-to-no external input other than that which an experimenter provides, hence the possibility for the network to become quiescent if it cannot sustain its activity through recurrent excitation.
The second type of transition occurs in spontaneously active networks, exhibiting a transition from asynchronous firing to high and low firing (Fig.~\ref{fig:introschematic}E-H).
This is appropriate as a model of \emph{in vivo} networks, neural tissue that is still part of the brain and receives input from other brain regions.
We will first show how to predict these transitions based on a mean-field analysis of the model, and then present the scaling theory that incorporates the effects of stochastic fluctuations that modify the universal quantitative properties that can be measured in experiments. 

\section{Widom scaling theory}
\label{subsec:widomtheory}

\subsection{General theory}
\label{sec:general}

As seen in the simulations in Fig.~\ref{fig:introschematic}, the spiking activity undergoes a phase transition at some critical synaptic coupling $J_c$ and critical baseline $\mathcal E_c$.
Because we assume our networks consist of statistically homogeneous neurons, we focus on the dynamics of the population averages of the neurons' membrane potentials and firing rates, $\psi(t) \equiv N^{-1} \sum_{i=1}^N \langle V_i(t)\rangle$ and $\nu(t) \equiv N^{-1} \sum_{i=1}^N \langle \dot{n}_i(t)\rangle$.
Away from the critical point, the population activity decays exponentially to a steady state, while at the critical point the activity decays algebraically---i.e., a power law.
However, we can make a stronger statement than this. 
For networks tuned close to the critical point $(\mathcal E_c,J_c)$ we anticipate that vestiges of scale invariance will lead to scaling collapse: although we are measuring the populations means $\psi(t)$ or $\nu(t)$ as a function of three independent parameters, time $t$, synaptic strength $J$, and baseline $\mathcal E$, for long times and small enough $J_c-J$ and $\mathcal E - \mathcal E_c$ we expect the data can be described by a function of only two combinations of the independent parameters.
The generic form of the ``Widom scaling relationship'' for the population-averaged firing rates is
\begin{align}
\nu(t) - \nu_c &\sim t^{-\frac{\beta_\ast}{\nu_\ast z_\ast}} F\left(|J_c-J|^{\nu_\ast z_\ast}t,(\mathcal E - \mathcal E_c)t^{\frac{\Delta_\ast}{\nu_\ast z_\ast}}\right),
\label{eqn:widomscalingform}
\end{align}
where $\nu_c$ is the firing rate at the critical point, $F$ is a scaling function of two arguments, and we have introduced the critical exponents $\beta_\ast$, $\nu_\ast$, $z_\ast$, and $\Delta_\ast$; we give critical exponents subscripts of $\ast$ to distinguish them from other variables that have similar symbols.
The exponents $\nu_\ast$, $z_\ast$, and $\eta_\ast$ are conventionally called the ``correlation length,'' ``dynamic,'' and ``anomalous'' exponents, while $\beta_\ast$ and $\Delta_\ast$ are derived from these exponents by scaling relations we will introduce later. 
We will retain the name ``correlation length exponent'' for $\nu_\ast$, even though there may not be a notion of length in arbitrary networks.

Note that the scaling function $F$ is actually multi-valued: it can depend on the sign of $J_c-J$, $\mathcal E - \mathcal E_c$, and $\nu(t)-\nu_c$.
This scaling form holds for both the \emph{in vitro} and \emph{in vivo} network models, though the values of the exponents and the scaling function $F$ will differ between the two cases because the two types of networks belong to different universality classes, as we will explain.
We derive this scaling form within both the mean-field approximation and our renormalization group analysis in Appendix~\ref{sec:scalingformderivation}.

Eq.~(\ref{eqn:widomscalingform}) tells us that if can determine the correct values of the critical exponents and the critical parameters $\nu_c$, $J_c$, and $\mathcal E_c$, then a plot of $(\nu(t)-\nu_c)t^{\beta_\ast/\nu_\ast z_\ast}$ against $|J_c-J|^{\nu_\ast z_\ast}t$ and $(\mathcal E - \mathcal E_c)t^{\Delta_\ast/\nu_\ast z_\ast}$, will ``collapse'' our three-dimensional dataset onto a two-dimensional surface.
In practice, such a collapse is difficult to achieve, and instead one tries to eliminate one of the variables by tuning it to its critical value, and then performing the collapse in the remaining variables, in which case the data should fall onto a one-dimensional curve.

In \emph{in vitro} networks we focus on the case $\mathcal E = \mathcal E_c$, and we will collapse firing rates $\nu(t)$ for different synaptic strengths using the reduced scaling form
\begin{equation}
\nu(t) \sim |J_c-J|^{\beta_\ast} F\left(|J_c-J|^{\nu_\ast z_\ast}t\right),
\label{eqn:ASwidomscaling}
\end{equation}
where $\nu_c = 0$ and we have pulled a factor of $(t |J_c-J|^{\nu_\ast z_\ast})^{\beta_\ast/\nu_\ast z_\ast}$ out of the scaling function $F$ in order to write the prefactor as $|J_c-J|^{\beta_\ast}$.

In our \emph{in vivo} model we will instead consider $J = J_c$, for which the scaling form can be reduced to
\begin{align*}
    \nu(t) - \nu_c &\sim t^{-\frac{\beta_\ast}{\nu_\ast z_\ast}}F\left((\mathcal E - \mathcal E_c)t^{\frac{\Delta_\ast}{\nu_\ast z_\ast}}\right).
\end{align*}
The practical difficulty with this scaling form is identifying the critical firing rate $\nu_c$.
This difficulty can be eliminated by performing two paired experiments: one in which the neurons are potentiated and then allowed to relax to the steady state from above, and another in which the neurons are suppressed and then allowed to relax to the steady state from below, with all other parameters being the same in the two experiments.
The scaling function $F$ will differ in these two scenarios (Appendix~\ref{sec:scalingformderivation}), allowing us to obtain a scaling form for the difference in activity:
\begin{align}
\nu_+(t) - \nu_-(t) \sim t^{-\frac{\beta_\ast}{\nu_\ast z_\ast}} \bar{F}\left((\mathcal E - \mathcal E_c)t^{\frac{\Delta_\ast}{\nu_\ast z_\ast}}\right),
\label{eqn:doublescalingform}
\end{align}
where $\bar{F} = F_+ - F_-$, where the sign subscripts correspond to potentiation ($+$) or suppression ($-$).

Next, we briefly review the mean-field approximation of the network activity and its predictions for the critical exponents, before highlighting the results of the renormalized scaling theory.

\subsection{Mean-field scaling theory}
\label{subsec:meanfield}

The stochastic system defined by Eqs.~(\ref{eqn:LIF})-(\ref{eqn:poiss}) cannot be solved in closed form, and understanding the statistical dynamics of these networks has historically been accomplished through simulations and approximate analytic or numerical calculations.
A qualitative picture of the dynamics of the model can often be obtained by a mean-field approximation in which fluctuations are neglected, such that $\langle \dot{n}_i(t) \rangle = \langle \phi(V_i(t)) \rangle \approx \phi\left(\langle V_i(t) \rangle\right)$, and solving the resulting deterministic dynamics: 
\begin{align}
\tau \frac{d \langle V_i(t) \rangle}{dt} &= -(\langle V_i(t) \rangle - \mathcal E) + \sum_{j=1}^N J_{ij} \phi\left(\langle V_j(t) \rangle\right).
\label{eqn:meanfielddynamics}
\end{align}
Equations of this form are a cornerstone of theoretical neuroscience \cite{SompolinskyPRL1988,vanVreeswijkScience1996,kadmon2015transition,marti2018correlations,schuecker2018optimal,stapmanns2020self}, though often motivated phenomenologically as firing rate models, rather than as the mean-field approximation of a spiking network's membrane potential dynamics. 
A wide variety of different types of dynamical behaviors and transitions among behaviors are possible depending on the properties of the connections $J_{ij}$ and nonlinearity $\phi(V)$ \cite{amari1977dynamics,rabinovich2006dynamical}, including bump attractors \cite{zhang1996representation,laing2001stationary,wimmer2014bump,kim2017ring}, pattern formation in networks of excitatory and inhibitory neurons \cite{ermentrout1979mathematical,bressloff2010metastable,butler2012evolutionary}, transitions to chaos \cite{SompolinskyPRL1988,dahmen2019second}, and avalanche dynamics \cite{beggs2003neuronal,buice2007field,FriedmanPRL2012}. 
In many of these examples, networks admit steady-states for which $d\langle V_i\rangle/dt = 0$ for all $i$ as $t \rightarrow \infty$.  

Within the mean-field approximation, the dynamics for the population-and-trial-averaged means reduce to
\begin{align}
\tau \frac{d \psi(t)}{dt} &= -(\psi(t) - \mathcal E) + J \Lambda_{\rm max} \nu(t),
\label{eqn:homogmodedynamics}
\end{align}
where $\nu(t) \approx \phi(\psi(t))$.
The phase transitions of the network can be characterized by analyzing the dynamics of this one-dimensional system.
While Eq.~(\ref{eqn:homogmodedynamics}) cannot be solved exactly, one can show that there is a continuous bifurcation at $(\mathcal E,J) = (\mathcal E_c,J_c)$, where the critical baseline $\mathcal E_c$ and critical synaptic weight $J_c$ are defined by
\begin{align}
    \mathcal E_c - \psi_c + J_c \Lambda_{\rm max} \phi(\psi_c) &= 0, \label{eqn:Ecrit} \\
    1 - J_c \Lambda_{\rm max}\phi'(\psi_c) &= 0,
\label{eqn:Jcrit}
\end{align}
in both the \emph{in vitro} and \emph{in vivo} network models.
The first condition, Eq.~(\ref{eqn:Ecrit}), corresponds to the baseline potential $\mathcal E$ and the mean synaptic input to each neuron balancing out so that the membrane potential of the neurons come to rest at $\psi_c$, to be determined momentarily.
The second condition, Eq.~(\ref{eqn:Jcrit}), corresponds to this steady state becoming marginally stable: perturbations away from the steady state will not decay away exponentially, nor will they grow exponentially.
We expect that when $J > J_c$ there will be multiple steady states, so in order for there to be a single marginal state the critical membrane potential $\psi_c$ must correspond to the largest value of the gain $\phi'(V)$,
\begin{align}
    \psi_c &= {\rm argmax}~\phi'(V). \label{eqn:psicrit}
\end{align}
In the \emph{in vivo} model this corresponds to $\psi_c = \theta$, while for the \emph{in vitro} model (with $r_0 > 1/2$), this corresponds to $\psi_c = \theta - \ln(r_0^{-1}-1)$, the activation threshold.
It follows that $\nu_c = 0$, $\mathcal E_c = \theta-\ln(r_0^{-1}-1)$, and $J_c = (\Lambda_{\rm max}r_0(1-r_0))^{-1}$ for the \emph{in vitro} model. 
For the \emph{in vivo} model we obtain $\nu_c = \phi(\theta)$, $\mathcal E_c = \theta - 2$, and $J_c = 4/\Lambda_{\rm max}$.

In our \emph{in vitro} model the phase transition separates an inactive steady state from an active steady state in which recurrent excitation is strong enough to self-sustain activity without external input, as shown in Fig.~\ref{fig:introschematic}A-D.
In our \emph{in vivo} model the phase transition separates an asynchronous steady state of intermediate firing rates from a bistable state of high- and low-firing rates, as shown in Fig.~\ref{fig:introschematic}E-H.

At long times and close enough to the critical point we expect $\psi(t)$ to be close to $\psi_c$, such that we can expand $\nu(t) = \phi(\psi(t)) = \sum_{\ell =0}^\infty \frac{\phi^{(\ell)}(\psi_c)}{\ell !} (\psi(t)-\psi_c)^\ell$ and truncate at the leading nonlinear order.
In our \emph{in vitro} model the leading nonlinear order is $\ell = 2$, while in the \emph{in vivo} model the leading order is $\ell = 3$.
Plugging this expansion into Eq.~(\ref{eqn:meanfielddynamics}) and solving for $\psi(t) - \psi_c$, and then approximating $\nu(t) \approx \phi(\psi_c) + \phi'(\psi_c)(\psi(t)-\psi_c)$ to leading order yields the scaling form (\ref{eqn:widomscalingform}) with exponents $\beta_\ast = 1$ and $\Delta_\ast = 2$ in the \emph{in vitro} models \footnote{There is an edge case that a reader might wonder about, $r_0 = 1/2$, for which $\phi''(\psi_c) = 0$ and the leading order nonlinearity is cubic, rather than quadratic. This represents a tricritical point of the system in which the mean-field exponents for the \emph{in vitro} model that match the exponents of the \emph{in vivo} model. In our simulations, which use the value of $r_0 = 1/2$, we observe the regular critical point because fluctuations induce a quadratic term in the effective nonlinearity. The tri-critical point, if it exists, requires a fine-tuned value of $r_0$ that we do not attempt to determine in this work.} and $\beta_\ast = 1/2$ and $\Delta_\ast = 3/2$ in the \emph{in vivo} model.
For both networks $\nu_\ast = 1/2$ and $z_\ast = 2$.

\begin{figure*}
 \centering
\includegraphics[scale=0.95]{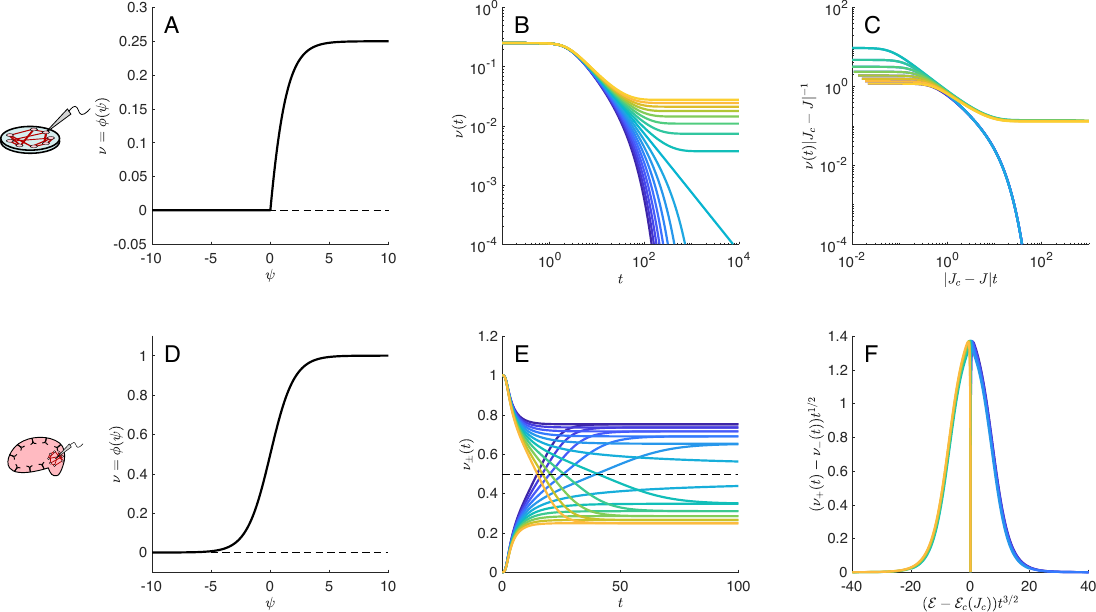}
  \caption{\textbf{Mean-field behavior of the spiking network} for \emph{in vitro} networks (top row) and \emph{in vivo} networks (bottom row). \textbf{A,D)} A typical nonlinearity $\nu = \phi(\psi)$ for each of the two network types. In the \emph{in vitro} networks the nonlinearity is rectified, such that the firing rate is zero when a neuron's membrane potential is negative. In \emph{in vivo} networks the firing rate is never zero---there is always a non-zero, though possibly small, probability of firing. \textbf{B,E)} The decay of $\nu(t)$, the population- and trial-averaged membrane potential, starting from an initial value of $\nu(0) \approx 1$ in \emph{in vitro} networks and $\nu_+(0) \approx 1$ and $\nu_-(0) \approx 0$ in \emph{in vivo} networks. \textbf{C,F)} Widom scaling collapses using Eq.~(\ref{eqn:ASwidomscaling}) for \emph{in vitro} networks and Eq.~(\ref{eqn:doublescalingform}) for \emph{in vivo} networks and the mean-field exponents given in Table~\ref{tab:latticeexponents}. }
  \label{fig:widomcollapse}
% \vspace{-.1in}
\end{figure*}

We verify that numerical solution of the full mean-field dynamics (Eq.~(\ref{eqn:meanfielddynamics})) indeed satisfies the scaling laws when the networks are tuned to their respective critical points.
We plot these collapses in Fig.~\ref{fig:widomcollapse} to illustrate the idea of data collapses.

\subsection{Renormalized scaling theory}
\label{sec:beyondmft}

While the mean-field approximation generally paints a qualitatively correct picture of phase transitions in a stochastic model, it is well-known that universal quantities like critical exponents or the Widom scaling functions $F$ are often quantitatively incorrect \cite{GoldenfeldBook1992}, a problem that was ultimately resolved by the development of the renormalization group \cite{kadanoff1966scaling,wilson1971renormalization,wilson1975renormalization,wilson1983renormalization}.

By developing an RG procedure that can be applied to this spiking network model (detailed in Appendix~\ref{sec:NPRG}), we can capture the effects of stochastic fluctuations on the collective activity of the network.
Within our RG approximation scheme, the dynamics of the population averages obey
\begin{align}
\tau \frac{d\psi(t)}{dt} &= -\psi(t) + \mathcal E + J\Lambda_{\rm max} \nu(t),
\label{eqn:truemeandyn}\\
\nu(t) &= \Phi(\psi(t)) \label{eqn:truenu},
\end{align} 
which is similar to Eq.~(\ref{eqn:meanfielddynamics}) except that the nonlinearity $\phi(\psi)$ is replaced with an effective nonlinearity $\Phi(\psi)$. 
The key idea behind the RG method presented in Appendix~\ref{sec:NPRG} is that we can compute this effective nonlinearity by iteratively averaging the bare nonlinearity over fluctuations associated with \emph{different eigenmodes} of the synaptic weight matrix $J_{ij}$.
In lattices these eigenmodes are simply Fourier bases, which can be parametrized in terms of spatial frequencies (``momenta'') which are traditionally coarse-grained in statistical physics. 
The eigenvalues of the synaptic weight matrix thus generalize the traditional ``momentum-shell'' RG approach, though it is not the only possible choice; see also \cite{bradde2017pca,meshulam2019coarse,kline2022gaussian,sooter2024cortex}.

Near the critical synaptic coupling this effective nonlinearity has the form
\begin{align}
\Phi(\psi) &= \nu_c + J^{-1}\Lambda_{\rm max}^{-1}(\psi - \psi_c) \label{eqn:Phiexpansion} \\
& ~~~~ + |J_c - J|^{\Delta_\ast} f^\ast((\psi-\psi_c)/|J_c-J|^{\beta_\ast}) + \dots, \nonumber 
\end{align}
where $\nu_c$ is the critical firing rate of a neuron, $\psi_c$ is the critical membrane potential, $f^\ast$ is a universal function, and $\beta_\ast$ and $\Delta_\ast$ are universal critical exponents.
Plugging (\ref{eqn:Phiexpansion}) into (\ref{eqn:truemeandyn})-(\ref{eqn:truenu}), the scaling forms (\ref{eqn:ASwidomscaling}) and (\ref{eqn:doublescalingform}) follow with non-trivial values of the critical exponents and scaling functions $F$. 

The scaling function and the critical exponents are characteristics of the ``universality class'' of a system, which is determined by the (emergent) symmetries of a model.
In lattice systems these universality classes are sub-divided by the spatial dimension of the system.
i.e., a two-dimensional system will have different critical exponents than a three-dimensional system, despite them both having the same underlying symmetries.

While several notions of dimension have been proposed for complex networks \cite{wen2021fractal}, it is not immediately clear which, if any, is the appropriate generalization in the context of critical phenomena.
In our RG analysis of the spiking network model (Appendix~\ref{sec:NPRG}), we find that the appropriate generalization of dimension is the \emph{spectral dimension} of the eigenvalue distribution $\rho_\lambda(\lambda)$ of the synaptic weight matrix $J_{ij}$, defined by
\begin{equation}
d/2-1 \equiv \lim_{\lambda \rightarrow \Lambda_{\rm max}^-} -(\Lambda_{\rm bulk} - \lambda) \frac{d}{d\lambda} \ln \rho_\lambda(\lambda),
\label{eqn:spectraldimdefn}
\end{equation}
where $\Lambda_{\rm bulk}$ is the maximum eigenvalue of the continuous part of the eigenvalue spectrum and $\Lambda_{\rm max}$ is the maximum eigenvalue of the network.
If $\Lambda_{\rm bulk} = \Lambda_{\rm max}$, then the spectrum is continuous near the maximum eigenvalue and $\rho_\lambda(\lambda) \sim (\Lambda_{\rm max} - \lambda)^{d/2-1};$
the definition of $d$ here is chosen so that it matches the spatial dimension when the network is a hypercubic lattice with periodic boundary conditions.
If $\Lambda_{\rm bulk} \neq \Lambda_{\rm max}$, then the largest eigenvalue is an outlier and $d$ diverges---this will be relevant in the case of random regular networks we study in Sec.~\ref{sec:applications}.
The spectral dimension has been identified as the relevant definition of dimension in other work investigating critical dynamics of, e.g., Ising-like models or random walks on networks \cite{tuncer2015spectral,millan2021complex,bighin2024universal}.

We can further show that the \emph{in vitro} and \emph{in vivo} models belong to different universality classes.
The \emph{in vitro} model belongs to the ``directed percolation'' universality class, a ubiquitous non-equilibrium universality class that describes the transition between extinction of activity and self-sustained activity.
The directed percolation universality class is characterized by an emergent ``rapidity symmetry'' that relates the magnitude of the membrane potential to fluctuations in the spiking activity, and three independent exponents, the correlation length exponent $\nu_\ast$, the dynamic exponent $z_\ast$, and the anomalous exponent $\eta_\ast$. 
The other critical exponents can be expressed as $\beta_\ast = \frac{\nu_\ast}{2}(d + \eta_\ast)$ and $\Delta_\ast = \frac{\nu_\ast}{2}(d + 2 z_\ast - \eta_\ast)$ \cite{janssen2005field}.
The directed percolation universality class has an ``upper critical dimension'' of $d = 4$, which means that networks with spectral dimension above this value will display mean-field scaling. 
Networks with $\Lambda_{\rm bulk} \neq \Lambda_{\rm max}$ have $d = \infty$, and are above the upper critical dimension.
The values of these critical exponents in lattice systems and mean-field are summarized in Table~\ref{tab:latticeexponents}.

Turning to the \emph{in vivo} model, our RG analysis predicts that the model belongs to the Ising model universality class, which describes transitions from a disordered state to ordered states related by an inversion symmetry.
In the spiking network the two ordered states are the high and low firing rate states, and the inversion symmetry implies that close to the phase transition the distribution of fluctuations around the means of the high and low firing rates are identical.
The universality class of the non-equilibrium Ising model is characterized by the correlation length exponent $\nu_\ast$, the dynamic exponent $z_\ast$, and anomalous exponent $\eta_\ast$, with $\beta_\ast = \frac{\nu_\ast}{2}(d-2+\eta_\ast)$ and $\Delta_\ast = \frac{\nu_\ast}{2}(d+2-\eta_\ast)$.
Like the directed percolation universality class, the Ising model has an upper critical dimension of $d = 4$, above which the mean-field approximation predicts the correct scaling.
The values of the critical exponents for two- and three-dimensional lattices and mean-field are given in Table~\ref{tab:latticeexponents}.

The values of the critical exponents are not necessarily the same for neurons arranged in lattices and complex networks, even if they have the same spectral dimension $d$.
Our RG scheme, described in Appendix~\ref{sec:NPRG}, does predict that lattices and networks with the same spectral dimension will have the same exponents, but we expect this to be only true approximately.
Being the first application of the RG to a spiking population model, our method does not capture the effects of the eigenmode structure of the synaptic weight matrix on the critical exponents, which could impact the values of $\eta_\ast$ and $z_\ast$ in particular, which our method predicts to have the mean-field values $\eta_\ast = 0$ and $z_\ast = 2$. 
However, our method does predict anomalous values of $\nu_\ast$, $\beta_\ast$, and $\Delta_\ast$. 
This said, because our RG scheme predicts the universality classes of the \emph{in vitro} and \emph{in vivo} networks, we can use the full set of anomalous exponents from $d$-dimensional lattices as a starting point for the scaling collapses we perform on our simulated data, allowing us to estimate potential discrepancies between lattices and networks with the same spectral dimension.

Finally, within our RG approximation we can also analytically compute the asymptotic tails of the scaling functions $F$ appearing in the Widom scaling forms.
For \emph{in vitro} networks we find
\begin{align}
F(x) &\sim \left\{\begin{array}{c c} \exp\left(-C_< x\right),&~J < J_c  \\ y_\infty + B_> \exp\left(-C_> x\right),&~J > J_c \end{array}\right., \label{eqn:FAStails}
\end{align}
where $x \propto |J_c-J|^{\nu_\ast z_\ast}t$ and the constants $C_<$, $C_>$, $B_>$, and $y_\infty$ are non-universal constants.
In \emph{in vivo} networks we find that the tails of the scaling function (\ref{eqn:doublescalingform}) obey
\begin{align}
\bar{F}(x) &\sim x^{\frac{\beta_\ast}{\Delta_\ast}} \exp\left(-C x^{1 - \frac{\beta_\ast}{\Delta_\ast} }\right), \label{eqn:Fbartails}
\end{align}
where $x \propto (\mathcal E - \mathcal E_c)t^{\frac{\Delta_\ast}{\nu_\ast z_\ast}}$ and $C$ is another non-universal constant.
Up to the non-universal constants, we show in the next section that not only can we collapse simulated activity data, the collapses agree well with the predicted scaling functions.

\section{Scaling analyses of simulated data}
\label{sec:applications}

To validate our scaling theory, we first show that simulated data from neurons arranged on $2$- and $3$-dimensional lattices with nearest-neighbor excitatory connections are indeed collapsed using the directed percolation or Ising model critical exponents listed in Table~\ref{tab:latticeexponents}.

We then investigate scaling in networks in which excitatory neurons are sparsely connected with fixed degree and inhibitory neurons, if present, provide broad global inhibition.
Our primary goal is to verify that the universality classes of the \emph{in vitro} and \emph{in vivo} models are consistent with the directed percolation and Ising universality classes, respectively.
We do not seek to obtain high precision estimates of the exponents competitive, instead devoting our computational resources to estimating the exponents on several network types.

\subsection{Excitatory lattices}

\begin{figure*}
 \centering
\includegraphics[scale=0.92]{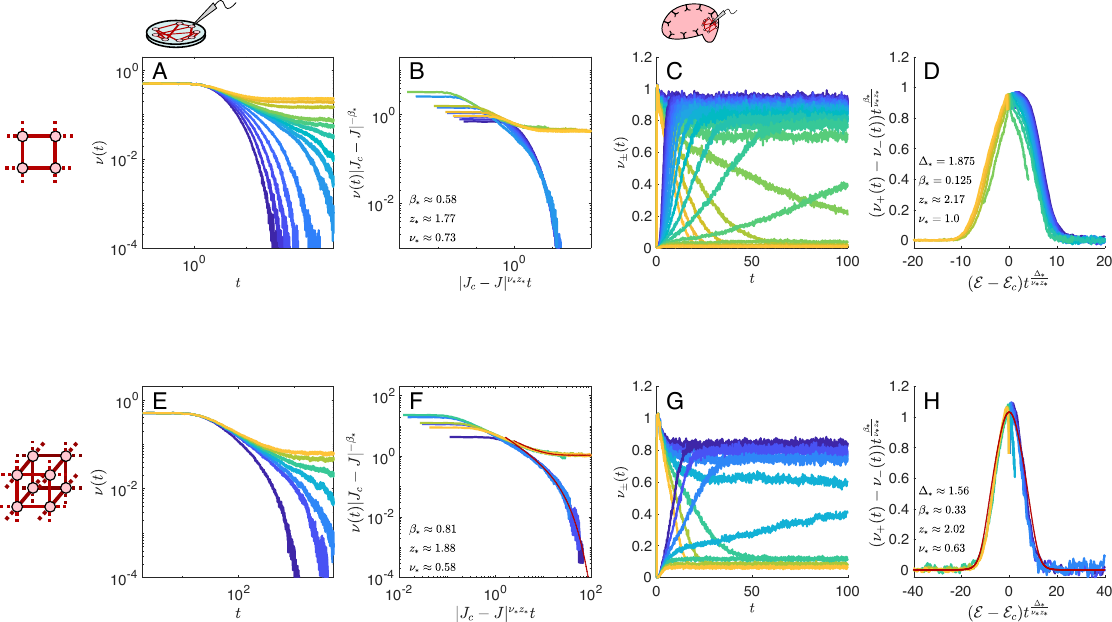}
  \caption{\textbf{Widom scaling collapses for simulated activity on excitatory lattices.} Top row (\textbf{A}-\textbf{D}): $d=2$. Bottom row (\textbf{E}-\textbf{H}): $d=3$. \textbf{A, E.}  Population-and-trial-averaged spike trains $\nu(t)$ versus time in \emph{in vitro} networks as the synaptic strength $J$ is tuned from subcritical ($J < J_c$, blue curves) to supercritical ($J > J_c$, green-gold curves). \textbf{B, F.} Widom scaling collapse of the data using Eq.~(\ref{eqn:ASwidomscaling}). Data below and above $J_c$ collapse onto different curves, with tails given by Eq.~(\ref{eqn:FAStails}). \textbf{C, G.} Population-and-trial-averaged spike trains versus time in \emph{in vivo} networks, starting from a high firing rate initial condition $\nu_+(0) \approx 1$ and a low firing rate initial condition $\nu_-(0) \approx 0$. Curves correspond to equilibrium potentials $\mathcal E > \mathcal E_c$ (green-gold curves) to $\mathcal E < \mathcal E_c$ (blue curves). \textbf{D, H.} Widom scaling collapse of $\nu_+(t)-\nu_-(t)$ according to Eq.~(\ref{eqn:doublescalingform}), with tails given by Eq.~(\ref{eqn:Fbartails}). The critical exponents used to collapse the data, inset in each collapse, are the known values of the critical exponents for the directed percolation (DP) and Ising model (IM) universality classes, given in Table~\ref{tab:latticeexponents}.}
  \label{fig:widomcollapse_latticesims}
% \vspace{-.1in}
\end{figure*}

As shown in Fig.~\ref{fig:widomcollapse_latticesims}A-B, simulated data from \emph{in vitro} models can be collapsed using the direction percolation critical exponents and the scaling form (\ref{eqn:ASwidomscaling}), and simulated data from \emph{in vivo} models can be collapsed using the Ising critical exponents and the scaling form (\ref{eqn:doublescalingform}). 

In our \emph{in vitro} networks the critical synaptic weights are $J_c \approx 1.56$ in $d=2$ and $J_c \approx 0.821$ in $d=3$, while in the spontaneous networks the critical parameters are $(\mathcal E_c,J_c) \approx (-4.27,3.3)$ in $d=2$ and $(\mathcal E_c,J_c) \approx (-2.79,1.165)$ in $d=3$.

\begin{table}[]
\begin{tabular}{cccc|ccc}
                                    & \multicolumn{3}{c|}{Directed percolation \cite{wang2013high}}                         & \multicolumn{3}{c}{Ising model}                                                        \\ \cline{2-7} 
\multicolumn{1}{c|}{}               & \multicolumn{1}{c|}{$d=2$}  & \multicolumn{1}{c|}{$d=3$}  & MF    & \multicolumn{1}{c|}{$d=2$ \cite{GoldenfeldBook1992}}  & \multicolumn{1}{c|}{$d=3$ \cite{kazmin2022critical}}  & \multicolumn{1}{c|}{MF}    \\ \hline
\multicolumn{1}{|c|}{$\nu_\ast$}    & \multicolumn{1}{c|}{$0.73$} & \multicolumn{1}{c|}{$0.58$} & $1/2$ & \multicolumn{1}{c|}{$1$}    & \multicolumn{1}{c|}{$0.63$} & \multicolumn{1}{c|}{$1/2$} \\
\multicolumn{1}{|c|}{$z_\ast$}      & \multicolumn{1}{c|}{$1.77$} & \multicolumn{1}{c|}{$1.89$} & $2$   & \multicolumn{1}{c|}{$2.17$ \cite{liu2023critical}} & \multicolumn{1}{c|}{$2.02$ \cite{adzhemyan2022dynamic}} & \multicolumn{1}{c|}{$2$}   \\
\multicolumn{1}{|c|}{$\eta_\ast$}      & \multicolumn{1}{c|}{$-0.41^\dag$} & \multicolumn{1}{c|}{$-0.17^\dag$} & $0$   & \multicolumn{1}{c|}{$1/4$} & \multicolumn{1}{c|}{$0.036$} & \multicolumn{1}{c|}{$0$}   \\
\multicolumn{1}{|c|}{$\beta_\ast$}  & \multicolumn{1}{c|}{$0.28$} & \multicolumn{1}{c|}{$0.82$} & $1$   & \multicolumn{1}{c|}{$1/8$}  & \multicolumn{1}{c|}{$0.33$} & \multicolumn{1}{c|}{$1/2$} \\
\multicolumn{1}{|c|}{$\Delta_\ast$} & \multicolumn{1}{c|}{$1.87^\dag$}       & \multicolumn{1}{c|}{$1.91^\dag$}       &   $2$    & \multicolumn{1}{c|}{$15/8$} & \multicolumn{1}{c|}{$1.56$} & \multicolumn{1}{c|}{$3/2$} \\ \hline
\end{tabular}
\caption{Critical exponents for the Directed percolation and Ising model universality classes on lattices of $d = 2$ and $3$ dimensions, compared to the mean-field (MF) prediction. Values are given to the hundredths place; more accurate estimates are available in the cited references. $^\dag$Derived exponents using the scaling relations $\beta^\ast = \frac{\nu_\ast}{2}(d+ \eta_\ast)$ and $\Delta_\ast = \frac{\nu_\ast}{2}(d+2z_\ast - \eta_\ast)$ for the directed percolation universality class \cite{janssen2005field}. }
\label{tab:latticeexponents}
\end{table}

\subsection{Sparse excitation and dense inhibition}
\label{sec:randreg}

We now consider networks with slightly more realistic features.
Cortical circuits consist of two broad cell types: excitatory and inhibitory.
Excitatory cells are typically thought to be the ``principal neurons'' whose activity is the neural realization of computations within cortical circuitry, while inhibitory cells are often ``interneurons'' that serve these computations indirectly by regulating the activity of the principal neurons.
In cortex, excitatory neurons have been found to make sparse connections to other excitatory neurons \cite{seeman2018sparse}, instead influencing each other through the densely connected inhibitory interneurons \cite{hofer2011differential}.
We therefore consider a network in which excitatory neurons make sparse connections to one another; specifically, we will model excitatory-excitatory connections using random regular graphs in which every neuron makes a fixed number of synaptic connections $k$ but the pairs of neurons connected are randomly chosen, independent of any spatial organization of the network.
The remaining connections in the network are dense; for simplicity we take the connections from excitatory to inhibitory cells, as well as inhibitory-to-inhibitory or inhibitory-to-excitatory, to be all-to-all connected.

\begin{figure*}
 \centering
\includegraphics[scale=0.95]{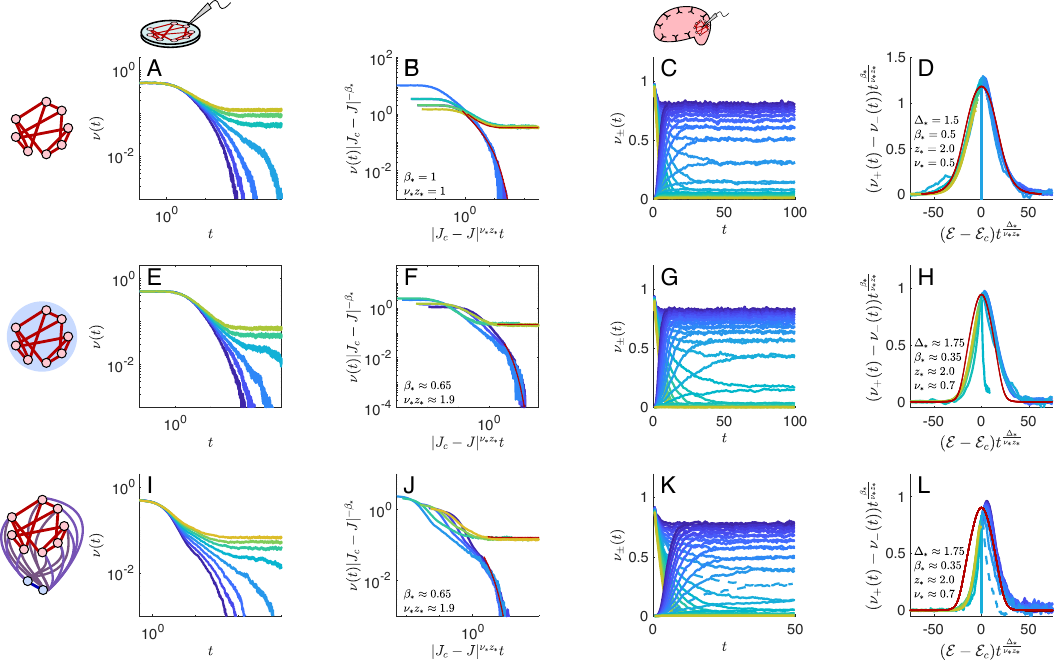}
  \caption{\textbf{Widom scaling collapses for simulated activity on networks with random regular excitatory-excitatory connections.} All excitatory neurons make $k=3$ excitatory connections to other neurons. Top row: Simulation results for purely excitatory networks. \textbf{A.} Decay of the population-averaged spiking activity in an absorbing state network for several values of coupling strength $J$, and \textbf{B.} its corresponding data collapse using mean-field predictions for the critical exponents. \textbf{C.} Decay of the population-averaged spiking activity in a spontaneously active network for several values of the input current $\mathcal E$, and \textbf{D.} its corresponding data collapse using mean-field predictions for the critical exponents.  Middle row: Simulation results for an excitatory population with effective inhibitory connections between neurons. \textbf{E-F} and \textbf{G-H} are the same as \textbf{A-B} and \textbf{C-D}, but using anomalous values of the critical exponents. Bottom: Simulation results for a model of separate excitatory and inhibitory populations that reduces to the effective model (Appendix~\ref{sec:EIreduction}). \textbf{I-J} and \textbf{K-L} are the same as \textbf{E-F} and \textbf{G-H}, using the same values of the anomalous exponents. In the absorbing state collapses (second column), the analytically estimated asymptotic Widom scaling forms Eq.~(\ref{eqn:FAStails}) are plotted in red, scaled by non-universal factors to match the data. Similarly for the spontaneous network collapses (fourth column) using Eqs.~(\ref{eqn:doublescalingform}) and (\ref{eqn:Fbartails}); see also Appendix~\ref{sec:scalingformderivation}. }
  \label{fig:widomcollapse_randreg_sims}
% \vspace{-.1in}
\end{figure*}

First, it is useful to consider what happens in networks without inhibition, for which we need only consider the excitatory neurons arranged in a random regular network.
If the synaptic strength of each connection is $J$ and each neuron has a refractory self-input of strength $-J$, then the eigenvalue distribution of the synaptic weight matrix has a maximum eigenvalue $J\Lambda_{\rm max} = J(k-1)$ associated with the homogeneous mode. 
This eigenvalue is an outlier.
The bulk spectrum of the synaptic weight matrix is given by the McKay law \cite{mckay1981expected} in the $N \rightarrow \infty$ limit (modified to include the self-coupling and normalized by $J$),
\begin{align*}
\rho_\lambda(\lambda) = \frac{2k}{\pi} \frac{\sqrt{4(k-1) - (\lambda+1)^2}}{4k^2 - 4(\lambda+1)^2};
\end{align*}
where $\lambda \in [-2\sqrt{k-1}-1,2\sqrt{k-1}-1]$.
Note that the bulk spectrum has spectral dimension $d = 3$, independent of the degree $k$.
The contribution of the single outlier eigenvalue contributes negligibly to the effective nonlinearity $\Phi(\psi)$, but it nonetheless controls the phase transition because it renders the spectral dimension to be $d = \infty$ (Eq.~(\ref{eqn:spectraldimdefn})).
We therefore expect that the excitatory random regular network will exhibit mean-field critical exponents; we confirm this in the scaling collapses of both \emph{in vitro} and \emph{in vivo} networks, shown in Fig.~\ref{fig:widomcollapse_randreg_sims}.
We find that the transitions occur approximately at $J_c \approx 2.17$ in \emph{in vitro} networks and $(\mathcal E_c, J_c) \approx (-3.5, 3.75)$ in \emph{in vivo} networks.

Next, we consider what happens when we turn on the inhibitory connections.
Rather than analyzing the full EI population, it is useful to first consider an \emph{effective} network model consisting of excitatory neurons that excite their random regular neighbors while inhibiting all other neurons in the network. 
This effective model is a formal reduction of a full population model with explicit excitatory and inhibitory populations; see Appendix~\ref{sec:EIreduction}.
Suppose the global inhibitory connections have strength $-J c/N$, where $N$ is the number of excitatory neurons.
These inhibitory connections will shift the location of $\Lambda_{\rm max}$ from $k-1$ to $k-1 - c$, without affecting any other eigenmodes of the network because they are orthogonal to the homogeneous mode.
There is then a critical value of $c = k-2\sqrt{k-1}$ for which the maximum eigenvalue is moved to the location of the bulk eigenvalue $\Lambda_{\rm bulk} = 2\sqrt{k-1}-1$, closing the gap between the bulk spectrum and the outlier.
We then expect the effective dimension to be $d = 3$, and the network may exhibit anomalous scaling instead of mean-field scaling.
We verify this for networks with $k=3$ for both the effective EI network and networks with explicit inhibitory neurons.

In the case of the effective EI networks, we find $J_c \approx 2.32$ with approximate exponents $\beta_\ast \approx 0.65$ and $\nu_\ast z_\ast \approx 1.9$ in \emph{in vitro} networks, and $J_c \approx 6.0$, $\beta_\ast \approx 0.35$, $\nu_\ast \approx 0.7$, $z_\ast \approx 2$ in \emph{in vivo} networks. 
The same exponents with $J_c \approx 2.39$ and $J_c \approx 6.0$ in \emph{in vitro} and \emph{in vivo} networks, respectively, produce collapses in simulations with explicit excitatory and inhibitory populations (Fig.~\ref{fig:widomcollapse_randreg_sims}).
While these estimates are not especially precise, the \emph{in vitro} exponents differ enough from the exponents in the $d = 3$ lattices to suggest that the network structure does have an influence on the critical exponents, and hence the universality class of the networks may differ from the $3$-dimensional lattices, although these universality classes are still in some sense close.

We see that the collapses for the effective and explicit EI networks are asymmetrically skewed compared to the analytically predicted Widom scaling forms for the $d=3$ Ising universality class, which match the $d=3$ lattice well.
Similar asymmetries have been observed in the scaling collapses of mean neural avalanche shapes, and recent modeling work has shown that inhibitory neurons play a role in producing these asymmetries \cite{zaccariello2025inhibitory}.
It is not clear what the origin of this asymmetry is.
Potentially it is a finite-size effect that will weaken in simulations of much larger networks. 
A surface plot of the firing rates versus $J$ and $\mathcal E$ does not show a sharp jump, compared to the lattices, suggesting that finite size effects may be softening the transition.
For instance, the dashed curve in Fig.~\ref{fig:widomcollapse_randreg_sims}K-L corresponds to a value of $\mathcal E$ that is close to the estimated critical $\mathcal E_c$, and does not collapse well in our current simulations, but may be more sharply separated from the critical point in larger networks.
Alternately, the asymmetries could be driven by subleading corrections to scaling or contributions from the network eigenmodes that are not captured by our current RG scheme.

Next, we perform a brief investigation of networks with degree $k > 3$.
There is a common folk wisdom that mean-field theory becomes accurate in high dimensional lattices because of the increased number of neighbors each unit interacts with. 
However, the spectral dimension of random regular networks does not depend on the number of neighbors $k$. 
We might therefore wonder whether the critical exponents will remain close to the values predicted by our scaling theory, or if at a sufficiently large $k$ we see a reversion to mean-field behavior.

We find that networks with $k \geq 4$ and global inhibition strength $c = k-2\sqrt{k-1}$ exhibit mean-field scaling.
However, the observed critical strength $J_c$ is lower than the mean-field prediction, in contrast to our other simulations.
This suggests that the global inhibition may be too large.
The mismatch most likely originates in our RG scheme's independence of the eigenmode structure of the network, and a higher order approximation scheme is required to identify the precise value of global inhibition at which anomalous scaling is observed for $k > 3$.

We may therefore wonder how robust the anomlous scaling of the $k=3$ network is to perturbations in the number of connections each neuron makes.
So far, we have considered networks in which the number of connections each neuron makes is the same for all neurons.
If we instead consider networks with a fraction $f$ of neurons that make $k=3$ synaptic connections and a fraction $1-f$ of neurons that make $k=4$ synaptic connections, we can investigate between which fractions we observe a transition from anomalous to mean-field scaling.

The excitatory synaptic connections between neurons are $J_{ij} = J\left(A_{ij} - (k_i-2) \delta_{ij} - c/N\right)$, where $A_{ij}$ is the adjacency matrix of the excitatory connections, formed by randomly paring up synapses of $fN$ degree $3$ neurons and $(1-f)N$ degree 4 neurons (and rejecting networks with self-connections or multiple synapses between a single pair of neurons); i.e., this is a configuration model \cite{fosdick2018configuring}. 
$k_i$ is the degree of neuron $i$, and $c$ is again the all-to-all inhibitory weight chosen to move the leading eigenvalue of $A_{ij} - (k_i-2) \delta_{ij}$ to the edge of the bulk spectrum.
Although we do not have a closed form for the spectrum of this weight matrix, the limiting cases $f = 1$ and $f = 0$---random regular networks of degrees $3$ and $4$, respectively---have spectral dimension $d=3$, and so we expect the mixed network does as well. 
We find that at $1-f = 10\%$ neurons with degree $4$ the anomalous scaling persists, while at $1-f = 20\%$ we again obtain mean-field scaling. 

\begin{figure*}
 \centering
\includegraphics[scale=1.1]{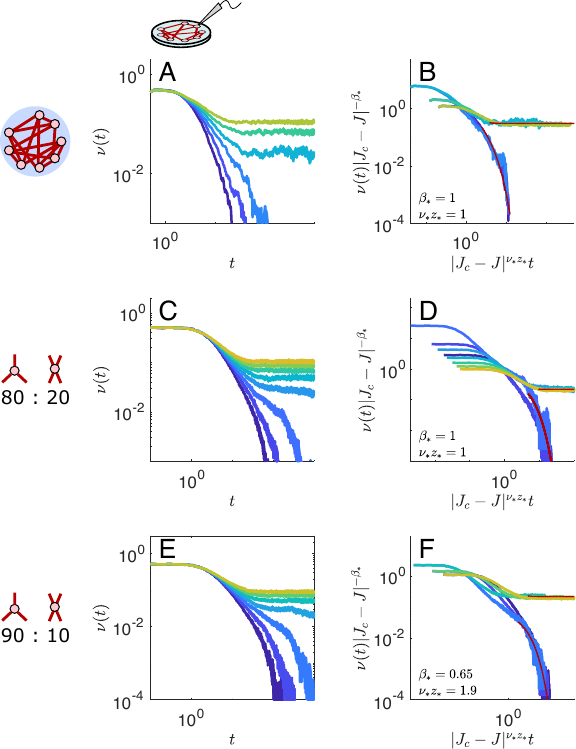}
  \caption{\textbf{Mean-field versus anomalous scaling in Excitatory-Inhibitory networks of varying degree.} \textbf{A-B.} Simulated activity in an effective EI network with degree $4$ random regular excitatory-excitatory connections, and \textbf{B.} its corresponding scaling collapse using mean-field exponents. The phase transition occurs at a value of $J_c \approx 1.38$ that is less than the mean-field prediction $J_c = 4/(2\sqrt{3}-1) \approx 1.62$, whereas in our other cases our RG analysis predicts that $J_c$ is larger than the mean-field prediction.  \textbf{C-D.} Effective EI network with $20\%$ degree $4$ and $80\%$ degree $3$ excitatory-excitatory connections. The phase transition occurs at $J_c \approx 1.98$, less than the mean-field prediction $J_c \approx 2.26$. The data can be collapsed using mean-field exponents. \textbf{E-F.} Effective EI network with $10\%$ degree $4$ and $90\%$ degree $3$ excitatory-excitatory connections. The estimated critical coupling is $J_c \approx 2.2$, comparable to the mean-field prediction of $J_c \approx 2.23$. The data, however, collapses using the same anomalous exponents as the pure degree $3$ effective EI network.}
  \label{fig:widomcollapse_randconfig_sims}
% \vspace{-.1in}
\end{figure*}

While we have not ruled out that there is a different value of global inhibition $c$ that can achieve anomalous scaling in networks with mixed connectivity, our results suggest a possible new mechanism that could explain apparently contradictory observations of mean-field versus anomalous scaling in neural avalanche data \cite{FriedmanPRL2012,yaghoubi2018neuronal,fontenele2019criticality}, with the observed scaling depending on the balance of excitatory sparsity and broad inhibition.

%%%%%%%%%%%%%%%%%%%%%%%%%%%%%%%
% DISCUSSION 
%%%%%%%%%%%%%%%%%%%%%%%%%%%%%%%

\section{Discussion}
\label{sec:discussion}

In this work we have shown that stochastic spiking networks with symmetric connections and homogeneous steady states can undergo at least two types of phase transitions as the strength of their synaptic connections $J$ and baseline potentials $\mathcal E$ are tuned: i) an inactive-to-active transition between extinction and self-sustained activity, appropriate as an \emph{in vitro} model of neural tissue, and ii) a transition from a single asynchronous state of activity to high or low firing rate states, appropriate as a model of \emph{in vivo} neural tissue.

Using both a mean-field approximation and a renormalization group analysis of the spiking network model, we developed a Widom scaling theory to show that the \emph{in vitro} network models belong to the directed percolation universality class, while the \emph{in vivo} network models belong to the Ising universality class.
These universality classes are subdivided by ``dimension,'' which we identified as the spectral dimension of the synaptic weight matrix.
If the largest eigenvalue of this spectrum is an outlier, then the spectral dimension is infinite, and we find the mean-field predictions of the critical exponents are correct.
However, if the largest eigenvalue is at the edge of the continuous part of the spectrum, then $d$ is finite and anomalous scaling may be observed.

This is, to the author's knowledge, the first renormalization group analysis of a leaky integrate-and-fire model.
While previous work modeling phase transitions in neural populations have used Ising-like models or chemical reaction networks of active or inactive neurons, these are phenomenological models of neural activity.
Similarly, while other work has investigated the non-perturbative renormalization group in neuroscience contexts, it has been applied only to calculating correlation and response functions in firing rate models \cite{stapmanns2020self}, and exploring possible equivalences between the RG and neural sensory coding work \cite{kline2022gaussian}.
This work establishes formally that spiking populations are in the Ising model or directed percolation universality classes.

The value of performing renormalization group calculations on spiking network models is that these calculations help clarify what features of neurons and their connectivity shape the critical exponents measured in data.
Experimental recordings of neural avalanches---cascades of neural activity triggered that propagate through neural circuitry \cite{beggs2003neuronal,BeggsFrontPhysio2012,buice2007field,FriedmanPRL2012}---often seem to support mean-field exponents, but deviations have also been reported \cite{FriedmanPRL2012}, and the origin of these deviations remains the subject of much debate.
Some reports suggest deviations could be the effects of subsampled recordings of neurons in space or time \cite{nonnenmacher2017signatures,levina2017subsampling}, while RG analysis of firing rate models suggest the cause could also be logarithmic corrections to critical exponents in networks at their upper critical dimension \cite{tiberi2022gell}.
Similarly, the ``phenomenological RG'' method developed by Ref.~\cite{meshulam2019coarse} motivates exponent relations based on analogies with lattice systems, and find anomalous scaling for the dynamical exponents $z_\ast$, albeit with values on the order of $0.16\sim0.3$, much smaller than $z_\ast \approx 2$ predicted by the directed percolation or Ising model universality classes.
Similar exponent values are obtained by Ref.~\cite{castro2024and} in rat visual cortex.
Using our foundational RG theory of spiking networks to understand the phenomenolgoical RG method could yield insight into these unexpectedly small dynamical exponents.

In general, having a firm theoretical understanding of what properties influence critical exponents will not only aid in disambiguating genuine deviations from mean-field theory versus estimates are skewed by subsampling problems, but could also reveal mechanisms by which a system that appears to be in a mean-field universality class could be tuned toward anomalous behavior.

For instance, both the directed percolation and Ising model mean-field universality classes make the same predictions for avalanche exponents, but in lower dimensions the exponents and even the exponent relations differ.
In our excitatory-inhibitory network of sparse excitatory connections that interact through dense inhibitory connections, we found that the inhibition could tune the network to a critical point with anomalous exponents for both the directed percolation and Ising universality classes.
One could imagine a future closed-loop experimental paradigm in which excitatory neurons are inhibited with wide-field optogenetic stimulation, where the strength of that stimulus depends on the recorded neural activity.
Depending on the properties of the excitatory connections (e.g., the degree of sparsity), this could drive a transition toward a anomalous critical state, distinguishing between different mean-field universality classes.

Several other mechanisms have been proposed as possible causes of mean-field or anomalous scaling. Ref.~\cite{tiberi2022gell} introduced a firing rate model in which nonlinearities in the membrane dynamics cancel out nonlinearities in the mean synaptic input, resulting in a new universality class with upper critical dimension $d = 2$. 
At this upper critical dimension the critical exponents differ from the mean-field scaling by logarithmic corrections that can depend on the distance to the critical point, introducing apparent anomalous scaling. 

Another mechanism well-known to change critical exponents is heterogeneity in network properties, often dubbed ``disorder'' in the statistical physics literature.
In this work we considered only heterogeneity in the pairs of neurons connected, and observed that it potentially alters critical exponents compared to lattice networks with the same effective dimension.
We assume our networks can be interpreted as the average connectivity in networks with weak variability that can be neglected.
Strong heterogeneity in the synaptic weights, however, can have a variety of possible effects.
It can smear out a transition, effectively destroying it,  drive the system to a different ``strong disorder'' universality class with different critical exponents, or  lead to anomalously slow temporal scaling $\exp(-t^a)$ or $t^{-b}$ for non-universal exponents $a$ and $b$ due to the existence of ``rare regions'' of the network that happen to be close to the critical point of the non-disordered system \cite{randeria1985low,bray1987nature,munoz2010griffiths}.

Because we have shown that the spiking network belongs to the directed percolation or Ising universality classes, we can leverage past work on other systems in the directed percolation \cite{hinrichsen1999model,hinrichsen2000possible,hooyberghs2004absorbing} or Ising classes \cite{vives1994avalanches,dahmen1996hysteresis} to motivate hypotheses for how heterogeneity might impact criticality in spiking networks.
For example, Ising models with heterogeneity in properties analogous to the baseline potential $\mathcal E$ and synaptic strength $J$ are thought to belong to the random field Ising model universality class \cite{vives1994avalanches,dahmen1996hysteresis}, for which renormalized scaling theories have been derived using extensions of the methods we use to obtain the results reported in this work \cite{tarjus2004nonperturbative,tarjus2008nonperturbative1,tissier2008nonperturbative2,balog2015activated,balog2018criticality}.

Heterogeneity in synaptic weights drives perhaps the most well known example of a phase transition in theoretical neuroscience, the celebrated transition to chaos in the Sompolinsky-Crisanti-Sommers model \cite{SompolinskyPRL1988,vanVreeswijkScience1996}, a firing rate model with random recurrent synaptic connections that can be interpreted as the mean-field theory of the spiking network studied in this work. 
This model and its many descendants have been extensively studied using methods from dynamic mean-field theory, and have become a cornerstone of theoretical neuroscience.
The transition to chaos is not in the directed percolation or Ising model universality classes, owing in part to the fact that the synaptic connections have zero mean and are not symmetric.
An important direction of future work is to extend our RG scaling theory to such networks and investigate whether stochastic spiking generates anomalous scaling in the transition to chaos.
While fluctuations have been added to this family of models by adding fluctuating external currents or Poisson inputs with rates matching the firing rate of the networks, mimicking the effect of spike fluctuations \cite{brunel2000dynamics,grytskyy2013unified}, and suggest that mean-field scaling persists in the presence of fluctuations, it remains an open question whether this is robust to adding structure to the synaptic connections, such as occurs in the networks studied in this work.

This said, all cases we have considered so far assume fixed synaptic connectivity. 
This is a reasonable assumption on timescales comparable to the millisecond scales of neural activity.
However, on longer timescales these connections can be modified by synaptic plasticity, in which neural activity drives strengthening or weakening of synaptic connections. 
Including synaptic plasticity in the stochastic spiking model couples the neural dynamics with their synaptic tuning parameters. 
Ref.~\cite{clark2024theory} investigated plasticity dynamics in a Sompolinsky-Crisanti-Sommers firing rate model using a dynamic mean-field approach, but this phenomena has yet to be explored through using RG scaling analyses.
It is often hypothesized that synaptic plasticity and homeostasis will lead to \emph{self-organized criticality}, with some simulation results supporting the possibility in simple models \cite{michiels2016synaptic}. 
This would imply that the synaptic strengths and baseline potentials of the networks are self-regulated towards their critical values over time \cite{vespignani1995renormalization,loreto1995renormalization,loreto1997loreto}.
Understanding the impact of disorder and synaptic plasticity on critical properties is crucial for interpreting neural data in the context of criticality, and the approach presented here establishes an important first step toward analyzing critical phenomena in spiking network models with heterogeneous features.

\section*{Acknowledgments}

The author thanks National Institute of Mental Health and National Institute for Neurological Disorders and Stroke grant UF-1NS115779-01 and Stony Brook University for financial support for this work, and Ari Pakman for feedback on an early version of this manuscript and anonymous referees for valuable feedback.

\appendix

%%%%%%%%%%%%%%%%%%%%%%%%%%%%%%%%
%% MODEL 
%%%%%%%%%%%%%%%%%%%%%%%%%%%%%%%%

\section{Derivation of the Widom scaling forms}
\label{sec:scalingformderivation}

Here we derive the Widom scaling forms (\ref{eqn:ASwidomscaling}) and (\ref{eqn:doublescalingform}), both in the mean-field approximation and using the results of our renormalization group analysis (described in Appendix~\ref{sec:NPRG}).

\subsection*{Mean-field scaling forms}

\subsubsection*{\emph{In vitro} networks}

In our \emph{in vitro} network models we set the baseline to its critical value, which we may take to be $\mathcal E_c = 0$, so we need only derive the scaling form as the synaptic weight $J$ is changed.
We choose parameter conventions so that the steady-state value of the membrane potential is $\psi_c = 0$, and we assume that $\phi''(0^+) < 0$.
Then, close to $\psi(t) = 0$ the mean-field dynamics (\ref{eqn:meanfielddynamics}) reduce to
\begin{align*}
    \tau \frac{d\psi(t)}{dt} &= -(1 - J\Lambda_{\rm max}  \phi'(0))\psi(t) \\
    & ~~~~~~~~~~~~ + \frac{J\Lambda_{\rm max}}{2!} \phi''(0)\psi(t)^2 + \dots.
\end{align*}
Neglecting the higher order terms, we can solve this by separation of variables, giving the implicit solution
\begin{align*}
\int_{\psi(0)}^{\psi(t)} \frac{dy}{-(1 - J\Lambda_{\rm max}  \phi'(0))y + \frac{J\Lambda_{\rm max}}{2!} \phi''(0)y^2} &= \frac{t}{\tau}.
\end{align*}
Although this integral can be evaluated exactly, to derive the scaling form it is useful to work with the integral form.
Comparing to Eq.~(\ref{eqn:ASwidomscaling}), we want a relationship between $(J_c-J)\psi(t)$ and $(J_c-J)t$, where $J_c - J \propto 1 - J\Lambda_{\rm max}\phi'(0)$.
This motivates a change of variables $\hat{y} =  y \xi_\tau$, where $\xi_\tau^{-1} \equiv (1 - J\Lambda_{\rm max}\phi'(0))/\tau$, which gives
\begin{align*}
\int_{\psi(0) \xi_\tau}^{\psi(t) \xi_\tau} \frac{d\hat{y}}{- \hat{y} + \frac{J\Lambda_{\rm max}}{2 \tau} \phi''(0) \hat{y}^2} &= \frac{t}{\xi_\tau}.
\end{align*}
If we assume that $\psi(0)\xi_\tau \gg 1$, such that we can approximate the lower limit as $\infty$ if $\xi_\tau > 0$ or $-\infty$ if $\xi_\tau < 0$, then the left hand side is a function of $\psi(t)\xi_\tau$ and the right-hand-side is $t/\xi_\tau$. 
Assuming that this function can be inverted gives a functional relationship that satisfies the scaling form (\ref{eqn:ASwidomscaling}).

We can check this against the exact solution,
$$\psi(t) = \frac{\psi(0) e^{-t/\xi_\tau}}{1 + \frac{J\Lambda_{\rm max} \xi_\tau |\phi''(0)|}{2 \tau} \psi(0) (1-e^{-t/\xi_{\tau}})},$$
If $0 < \xi_\tau < \infty$ and $\xi_\tau \psi(0) \gg 1$, then for long times this reduces to $\psi(t) \approx \frac{2 \tau}{J\Lambda_{\rm max}  |\phi''(0)|} \xi_\tau^{-1}  \exp(-t/\xi_{\tau})$, which reveals that $F(x) \propto \exp(-x)$.

If $-\infty < \xi_\tau < 0$, then for long times $\psi(t) \approx \frac{2 \tau}{J\Lambda_{\rm max}  |\phi''(0)|} |\xi_\tau^{-1}|  (1-\exp(-t/|\xi_{\tau}|))$, giving $F(x) \propto 1 - \exp(-x)$.

Finally, if $\xi_\tau \rightarrow \infty$ we find $\psi(t) \approx \frac{2\tau}{J_c\Lambda_{\rm max}|\phi''(0)|}t^{-1}$, the expected power-law decay. 

\subsubsection*{\emph{In vivo} networks}

We will work at the critical synaptic weight $J = J_c$, where $J_c = (\Lambda_{\rm max}\phi'(\psi_c))^{-1}$, where $\psi_c = \theta$ is the potential for which $\phi''(V) = 0$.
We expand the nonlinearity in the mean-field equation (\ref{eqn:meanfielddynamics}) around $\theta$, yielding
\begin{align*}
    \tau \frac{d}{dt}(\psi(t)-\theta) &= \mathcal E - \mathcal E_c \\
    &+ \frac{J\Lambda_{\rm max}}{3!} \phi^{(3)}(\theta)(\psi(t)-\theta)^3 + \dots
\end{align*}
where the linear term has vanished because $J = J_c$ and we defined $\mathcal E_c \equiv \theta - J_c\Lambda_{\rm max} \phi(\theta)$. 
Solving the equation implicitly, with $u(t) = \psi(t)-\theta$,
\begin{align*}
    \int_{u(0)}^{u(t)} \frac{dy}{\mathcal E - \mathcal E_c + \frac{J\Lambda_{\rm max}}{3!}\phi^{(3)}(\theta) \hat{y}^3} &= \frac{t}{\tau}
\end{align*}
In the mean-field approximation we expect the scaling variables to be $(\psi(t)-\theta)t^{1/2}$ and $(\mathcal E - \mathcal E_c)t^{3/2}$.
This motivates a change of variables $\hat{y} = y (t/\tau)^{1/2}$, which gives
\begin{align*}
    \int_{u(0)(t/\tau)^{1/2}}^{u(t)(t/\tau)^{1/2}} \frac{d\hat{y}}{(\mathcal E - \mathcal E_c)(t/\tau)^{3/2} - \frac{J\Lambda_{\rm max}}{3!}|\phi^{(3)}(\theta)| \hat{y}^3} &= 1
\end{align*}
Assuming that $|u(0)|(t/\tau)^{1/2} \gg 1$, we can approximate the lower limit as $-\infty$ if $u(0) < 0$ and $+\infty$ if $u(0) > 0$.
Note that although the value of $\psi(0)$ will not contribute to the asymptotic scaling form, its sign does determine the relevant branch of the evaluation of the integral, which will yield two different scaling forms.
The left-hand-side is then a function of both $(\psi(t)-\theta)(t/\tau)^{1/2}$ and $(\mathcal E - \mathcal E_c)(t/\tau)^{3/2}$, which is equal to a constant, imposing a functional relationship between the two scaling variables.
The scaling forms will both be of the form $(\psi(t) - \theta)(t/\tau)^{1/2} = F_\pm((\mathcal E - \mathcal E_c)(t/\tau)^{3/2})$.
Once we obtain the scaling forms numerically, we can subtract them to get a single scaling form for $\psi_+(t) - \psi_-(t)$, which will be of the form (\ref{eqn:doublescalingform})

\subsection*{Renormalized scaling theory}

The derivations of the Widom scaling forms using the effective nonlinearity $\Phi(\psi)$ are similar to the mean-field calculations, with some additional subtleties to manage.

Plugging the expansion of $\Phi(\psi)$, Eq.~(\ref{eqn:Phiexpansion}), into the homogeneous dynamics of the trial-averaged means, Eqs.~(\ref{eqn:truemeandyn})-(\ref{eqn:truenu}), and keeping only the leading order nonlinear contribution to the effective nonlinearity:
\begin{align*}
    \tau \frac{d\psi}{dt} &= -\psi + \mathcal E + J\Lambda_{\rm max} \Phi(\psi)\\
    &= \mathcal E - \mathcal E_c \\
    & ~~~~~~~ + J\Lambda_{\rm max}|J_c-J|^{\Delta_\ast}\\
    & ~~~~~~~~~~~ \times f^\ast\left((\psi-\psi_c)|J_c-J|^{-\beta_\ast}\right);
\end{align*}
where we define $\mathcal E_c \equiv \psi_c + \Lambda_{\rm max}\nu_c$.
Next, we consider our two specific network cases.

\subsubsection*{\emph{In vitro} networks}

For the \emph{in vitro} network models we set $\mathcal E = \mathcal E_c = 0$ and may take $\psi_c = 0$, so we need to evaluate the integral
\begin{align*}
\int_{\psi(0)}^{\psi(t)} \frac{ds}{|J_c-J|^{\Delta_\ast} f^\ast\left(s |J_c-J|^{-\beta_\ast}\right)} & = J_c\Lambda_{\rm max}\frac{t}{\tau}.
\end{align*}
In Appendix~\ref{sec:NPRG} we show that the scaling function in the denominator may be written $f^\ast(z) = c v_1(z) + \varphi_\ast(z)$, where we have estimated both $v_1(z)$ and $\varphi_{1\ast}(z)$ to order $z^5$. 
The two functions are characteristics of the directed percolation universality class, where $\varphi_{1\ast}(z)$ is a dimensionless counterpart of the effective firing rate nonlinearity $\Phi(\psi)$ and $v_1(z)$ characterizes deviations from the critical point.
The constant $c$ weights the deviation $v_1(z)$ (really a ``relevant eigenmode'' of the RG fixed point) against $\varphi_{1\ast}(z)$.
The sign of this weight depends on the sign of $J_c-J$.
For $c < 0$ we expect that $f^\ast(z)$ has no zeros for all $z > 0$ and the integral will exist for all $0 \leq \psi(t) \leq \psi(0)$. 
For $c > 0$ $f^\ast(z)$ may vanish, such that the integral will only exist for $s$ greater than the largest root of $f^\ast(z)$.
We cannot predict the precise value of $c$ with the approximation scheme we use in this work, so we will choose a sufficiently large magnitude for which $f^\ast(z)$ switches from having no zeros when $c < 0$ to at least one zero when $c > 0$.

We will first derive the general scaling form (\ref{eqn:ASwidomscaling}).
We first make the change of variables $s' = s|J_c-J|^{-\beta_\ast}$ and move the factors of $|J_c-J|$ to the right-hand side, giving
\begin{align*}
\int_{y_0}^{y} \frac{ds'}{f^\ast\left(s'\right)} & = x,
\end{align*}
where $y = \psi(t) |J_c-J|^{-\beta_\ast}$, $y_0 = \psi(0)|J_c-J|^{-\beta_\ast}$, and $x = \Lambda_{\rm max}|J_c-J|^{\Delta_\ast-\beta_\ast} \frac{t}{\tau} = \Lambda_{\rm max}|J_c-J|^{\nu_\ast z_\ast} \frac{t}{\tau}$, where we use our RG scheme's predictions that $z_\ast = 2,~\beta_\ast = \frac{\nu_\ast d}{2},~\Delta_\ast = \frac{\nu_\ast}{2}(d+4)$ to reduce $\Delta_\ast - \beta_\ast = \nu_\ast z_\ast$.
Because we are interested in the asymptotic limit of $J \simeq J_c$, we will take $y_0$ to be large enough that we can replace the lower limit of the integral by $+\infty$.
Within this approximation, the left-hand-side is a function of $y$ and the right-hand-side is $x$, so assuming the function can be inverted we obtain the scaling form (\ref{eqn:ASwidomscaling}).

We can derive the asymptotic tails explicitly using the results of our renormalization group analysis, covered in Appendix~\ref{sec:NPRG}. 
We expand $f^\ast(z) \approx (f^\ast)'(z_0)(z-z_0) + \frac{1}{2} (f^\ast)''(z_0)(z-z_0)^2$, cutting the series off at quadratic order, where $z_0$ is the largest zero of $f^\ast(z)$.
Assuming $y \geq z_0$ we can evaluate the integral analytically, just as in the mean-field case.
We obtain the general form $y \sim F(x)$ with
\begin{align}
F(x) &= z_0 + \frac{2 (f^\ast)'(z_0)}{(f^\ast)''(z_0)} \frac{e^{(f^\ast)'(z_0)x}}{1-e^{(f^\ast)'(z_0)x}}. \label{eqn:ASwidomtails_appendix}
\end{align}
We expect $(f^\ast)'(z_0) < 0$ and $(f^\ast)''(z_0) < 0$, such that for large $x$ the scaling variable $y$ approaches $z_0$ exponentially from above.
For $c < 0$ and sufficiently large in magnitude we expect $z_0 = 0$, while for $c > 0$ and large enough we expect $z_0 > 0$.

We can approximate the constants in Eq.~(\ref{eqn:ASwidomtails_appendix}) using our $\mathcal O(z^5)$ approximation in $d=3$.
If we normalize the eigenmode $v_1(z)$ such that $\int_0^\infty dz~e^{-z} v_1(z)^2 = 1$, then $|c| = 1.5$ provides a sufficient dividing line between subcritical and supercritical behavior; the actual value could be larger.
We find that for $c = -1.5$, $z_0 = 0$ as expected, and $(f^\ast)'(z_0) = -0.017$ and $(f^\ast)''(z_0) = -0.75$.
For $c = +1.5$ the largest real root occurs at $z_0 = 1.80$ and $(f^\ast)'(z_0) = -0.74$ and $(f^\ast)''(z_0) = -0.93$.
The coefficients of $v_1(z)$ and $\varphi_1^\ast(z)$ are given in Table~\ref{tab:AScoeffs}.

Finally, because $\nu(t) = \Phi_1(\psi(t)) = \Lambda_{\rm max}^{-1}\psi(t) + |J_c-J|^{\Delta_\ast}f^\ast(\psi(t)|J_c-J|^{-\beta_\ast}) + \dots$, to leading order $\nu(t)|J_c-J|^{-\beta_\ast}$ obeys the same scaling form as $y = \psi(t)|J_c-J|^{-\beta_\ast}$, at least for $J$ close enough to $J_c$. 

\begin{table}[]
\begin{tabular}{c|c|c|}
\cline{2-3}
                            & $v_1(z)$ & $\varphi_1^\ast(z)$ \\ \hline
\multicolumn{1}{|c|}{$z$}   &     $0.22$     &           $0.31$           \\
\multicolumn{1}{|c|}{$z^2$} &     $0.067$     &          $-0.65$            \\
\multicolumn{1}{|c|}{$z^3$} &      $-0.033$    &           $-0.25$           \\
\multicolumn{1}{|c|}{$z^4$} &    $0.025$      &             $0.089$         \\
\multicolumn{1}{|c|}{$z^5$} &      $-0.0033$    &             $-0.033$         \\ \hline
\end{tabular}
\caption{Coefficients of the $\mathcal O(z^5)$ estimates of the absorbing state network's critical dimensionless nonlinearity $\varphi_1^\ast(z)$ and the relevant eigenmode $v_1(z)$ in $d=3$. i.e., the $n^{\rm th}$ row gives the value of $v_1^{(n)}(0)$ or $(\varphi_1^\ast)^{(n)}(0)$.}
\label{tab:AScoeffs}
\end{table}

\subsubsection*{\emph{in vivo} networks}

For \emph{in vivo} networks we focus on the limit $J \rightarrow J_c$, for which $|J_c-J|^\Delta_\ast f^\ast((\psi-\psi_c)/|J_c-J|^\beta_\ast) \rightarrow \mathcal A[\psi-\psi_c]^{\Delta_\ast/\beta_\ast}$, where $\mathcal A$ is a universal constant that depends on the spectral dimension $d$ and we define the notation $[\psi-\psi_c]^{\Delta_\ast/\beta_\ast} = {\rm sgn}(\psi-\psi_c)|\psi-\psi_c|^{\Delta_\ast/\beta_\ast}$.
We thus need to evaluate the integral
\begin{align*}
\int_{\psi(0)}^{\psi(t)} \frac{ds}{\mathcal E - \mathcal E_c - \Lambda_{\rm max}\mathcal A [s-\theta^\ast]^{\Delta_\ast/\beta_\ast}} & = \frac{t}{\tau}.
\end{align*}
Because we want to obtain the scaling function corresponding to the scaling form Eq.~(\ref{eqn:doublescalingform}), we make the change of variables $s' = (s-\psi_c)(t/\tau)^{\frac{\beta_\ast}{\nu_\ast z_\ast}}$, giving
\begin{align*}
\int_{y_0}^{y} \frac{ds'}{x - \Lambda_{\rm max}\mathcal A [s']^{\Delta_\ast/\beta_\ast}} & = 1,
\end{align*}
where $y = (\psi(t)-\psi_c)(t/\tau)^{\frac{\beta_\ast}{\nu_\ast z_\ast}}$, $y_0 = (\psi(0)-\psi_c)(t/\tau)^{\frac{\beta_\ast}{\nu_\ast z_\ast}}$ and $x = (\mathcal E - \mathcal E_c)(t/\tau)^{\frac{\Delta_\ast}{\nu_\ast z_\ast}}$, and we used the fact that within our RG scheme we have the approximate scaling relations $(\Delta_\ast-\beta_\ast)/\nu_\ast/z_\ast = (d+2-(d-2))/2 = 1$.
Because we are interested in the asymptotic scaling in the long-time limit $t \gg \tau$, we can take the lower limit $y_0$ to be $\pm \infty$, depending on whether $\psi(0)$ is above or below $\theta^\ast$.
In addition to the sign of $y_0$, we also have to consider the sign of $x$ and the sign of $y$ to evaluate the integral.

If $x > 0$ and $y_0 > 0$, we expect that $y > 0$ for all time, and we need to evaluate the integral
\begin{align*}
-\int^{+\infty}_{y} \frac{ds}{x - \Lambda_{\rm max}\mathcal A s^{\Delta_\ast/\beta_\ast}} & = 1,
\end{align*}
which requires $x < \Lambda_{\rm max}\mathcal A y^{\Delta_\ast/\beta_\ast}$.
Similarly, if $x < 0$ and $y_0 < 0$, we expect $y < 0$ for all time and we need to evaluate
\begin{align*}
\int_{-\infty}^{y} \frac{ds}{-|x| + \Lambda_{\rm max}\mathcal A (-s)^{\Delta_\ast/\beta_\ast}} & = 1,
\end{align*}
which requires $-|x|  + \Lambda_{\rm max}\mathcal A (-y)^{\Delta_\ast/\beta_\ast} > 0$. 
A change of variables $s = -y$ transforms this integral to the $x > 0,~y > 0$ case.

The next two cases are slightly more complicated.
When $x > 0$ but $y_0 < 0$, then we expect $y$ to cross from initially negative values to positive values.
For $y < 0$ we may solve
\begin{align*}
\int_{-\infty}^{y} \frac{ds}{x + \Lambda_{\rm max}\mathcal A (-s)^{\Delta_\ast/\beta_\ast}} & = 1,
\end{align*}
while for $y > 0$ we need to split up the integral,
\begin{align*}
\int_{-\infty}^{0} \frac{ds}{x + \Lambda_{\rm max}\mathcal A (-s)^{\Delta_\ast/\beta_\ast}} + \int_{0}^{y} \frac{ds}{x - \Lambda_{\rm max}\mathcal A s^{\Delta_\ast/\beta_\ast}}  & = 1.
\end{align*}
Finally, for $x < 0$ but $y_0 > 0$ we expect $y$ to be initially positive and cross over to negative values, so we may write
\begin{align*}
-\int^{+\infty}_{y} \frac{ds}{-|x| - \Lambda_{\rm max} \mathcal A s^{\Delta_\ast/\beta_\ast}} & = 1,
\end{align*}
for $y > 0$ and 
\begin{align*}
1 &=-\int^{+\infty}_{0} \frac{ds}{-|x| - \Lambda_{\rm max}\mathcal A s^{\Delta_\ast/\beta_\ast}} \\
&~~~~~ - \int_{y}^{0} \frac{ds}{-|x| +  \Lambda_{\rm max}\mathcal A (-s)^{\Delta_\ast/\beta_\ast}}.
\end{align*}
for $y < 0$.
For $d=3$ and $d=4$ (mean-field) our RG scheme predicts integer values of $\Delta_\ast/\beta_\ast = (d+2)/(d-2)$, and the integrals can be evaluated analytically using Mathematica, though the result involves complex-valued representations that ultimately work out to be real and are otherwise not enlightening.
The implicit representation cannot be solved in closed form, but can be evaluated numerically to obtain scaling forms for $y_+$ and $y_-$, corresponding to the cases $y_0 > 0$ and $y_0 < 0$, respectively.
Subtracting these numerical solutions yields the scaling form for $y_+ - y_-$ plotted in Figs.~\ref{fig:widomcollapse_randreg_sims}D,H,L.

To estimate the asymptotic tails of the distribution for large $x > 0$, we make an additional change of variables $q' = \left(\frac{\Lambda_{\rm max}\mathcal A}{x}\right)^{\frac{\beta_\ast}{\Delta_\ast}}s'$.
For the $y_0 > 0$ case this gives
\begin{align*}
&\frac{1}{x}\left(\frac{\Lambda_{\rm max}\mathcal A}{x}\right)^{-\frac{\beta_\ast}{\Delta_\ast}}\int_q^\infty \frac{dq'}{(q')^{\Delta_\ast/\beta_\ast}-1}\\
 &= \frac{1}{x}\left(\frac{\Lambda_{\rm max}\mathcal A}{x}\right)^{-\frac{\beta_\ast}{\Delta_\ast}}\int_q^\infty \frac{dq'}{(q'-1) \frac{(q')^{\Delta_\ast/\beta_\ast}-1}{q'-1}} \\
&= \frac{1}{x}\left(\frac{\Lambda_{\rm max}\mathcal A}{x}\right)^{-\frac{\beta_\ast}{\Delta_\ast}}\left(\frac{\ln(q'-1)}{\frac{(q')^{\Delta_\ast/\beta_\ast}-1}{q'-1}}\Bigg|^{\infty}_q \right) \\
& ~~~~  - \int_q^\infty dq' ~\ln (q'-1) \frac{d}{dq'}\left({ \frac{(q')^{\Delta_\ast/\beta_\ast}-1}{q'-1}}\right)\\
&\approx -\frac{1}{x} \left(\frac{\Lambda_{\rm max}\mathcal A}{x}\right)^{-\frac{\beta_\ast}{\Delta_\ast}} \frac{\ln(q-1)}{\Delta_\ast/\beta_\ast} + \dots
\end{align*}
where in the last line we retain only the leading order behavior as $q \rightarrow 1$ and neglect the higher order terms from the integration by parts.
Solving for $q$ and writing the result in terms of $y_+$ gives
$$y_+ \sim 1 + \exp\left(-\frac{\Delta_\ast}{\beta_\ast}\left(\frac{\Lambda_{\rm max}\mathcal A}{x}\right)^{\frac{\beta_\ast}{\Delta_\ast}}x\right).$$

In the $y_0 < 0$ and $x > 0$ case we make a similar change of variables, and must evaluate
\begin{align*}
&\frac{1}{x} \left(\frac{\Lambda_{\rm max}\mathcal A}{x}\right)^{-\frac{\beta_\ast}{\Delta_\ast}} \\
& ~~~~ \times \left(\int_{-\infty}^0 \frac{dq'}{1-(-q')^{\Delta_\ast/\beta_\ast}} + \int_-^q \frac{dq'}{1-(q')^{\Delta_\ast/\beta_\ast}} \right).
\end{align*}
The first integral evaluates to a constant, $\frac{\pi}{\Delta_\ast/\beta_\ast} {\rm csc}\left(\frac{\pi}{\Delta_\ast/\beta_\ast}\right)$, while the second can be evaluated using a similar integration-by-parts trick. 
The result in terms of $y_-$ is
$$y_- \sim 1 - \exp\left(\pi {\rm csc}\left(\frac{\pi}{\Delta_\ast/\beta_\ast}\right) -\frac{\Delta_\ast}{\beta_\ast}\left(\frac{\Lambda_{\rm max}\mathcal A}{x}\right)^{\frac{\beta_\ast}{\Delta_\ast}}x \right).$$
Subtracting $y_+-y_-$ and omitting constant factors gives a scaling form for the difference of membrane potentials.
The $x < 0$ case may be obtained by replacing $x$ with $|x|$.
To obtain Eq.~(\ref{eqn:Fbartails}), which is a scaling form for the spike count differences, we use the fact that $\Phi(\psi(t)) = \nu_c + (J\Lambda_{\rm max})^{-1}(\psi(t) - \psi_c) + \dots$, so the leading order scaling is $\nu_+(t) - \nu_-(t) = (J_c\Lambda_{\rm max})^{-1}(\psi_+(t) - \psi_-(t)) + \dots$, giving Eq.~(\ref{eqn:Fbartails}).

\section{Reduction of an excitatory-inhibitory network to a random regular network with effective all-to-all inhibition}
\label{sec:EIreduction}

The random regular network with all-to-all inhibitory connections considered in Sec.~\ref{sec:randreg} is a reduction of a network model with explicit excitatory and inhibitory populations (an ``EI'' network).
In this network the E-E connections follow the random-regular network, while connections to and from the inhibitory population are all-to-all.
Here we present the heuristic derivation of this reduction, which is confirmed by the simulation results given in Fig.~\ref{fig:widomcollapse_randreg_sims}.

We follow the idea of Ref.~\cite{gerstner2014neuronal}, which performs a similar reduction of a mean-field model for an EI network. 
The idea behind the derivation is that we want the membrane timescale of the inhibitory population to be fast, such that the membrane potential of the inhibitory neurons very closely follows its input.
Second, we introduce synaptic gate dynamics for the inhibitory neurons with synaptic timescale long enough that the filtered spike trains approximately average the rate, allowing us to use a mean-field approximation for the inhibitory spikes.

The dynamics of the inhibitory neurons is given by
\begin{align*}
\tau_I \frac{dV_i(t)}{dt} &= -V_i + \mathcal E_I + \sum_{j \in E} J_{ij} \dot{n}_j(t) + \sum_{j \in I} J_{ij} s_j(t),\\
\tau^{\rm syn}_I \frac{ds_i}{dt} &=-s_i(t) + \dot{n}_i(t),\\
\dot{n}_i(t)dt &\sim {\rm Poiss}\left[\phi_I(V_i(t))dt\right],\\
\phi_I(V) &= \gamma\lfloor V_i - \theta_I \rfloor_+,
\end{align*}
where $\tau_I$ is the membrane time constant, $\mathcal E_I$ is the tonic input to each inhibitory cell, $\sum_{j \in E} J_{ij} \dot{n}_j(t)$ is the input from excitatory neurons, and $\sum_{j \in I} J_{ij} s_j(t)$ is the input from other inhibitory neurons.
The variables $s_i(t)$ are the synaptic gating variables of the inhibitory neurons, which essentially amount to an exponential filtering of the inhibitory spike trains with time constant $\tau^{\rm syn}_I$. 
The spikes are again conditionally Poisson with rate $\phi_I(V)$, which we take to be rectified linear with a gain $\gamma$ and threshold $\theta_I$, regardless of whether the E population is an \emph{in vitro} or \emph{in vivo} network.
The choice of rectified linear is so that the rate will be linear if the input is above the set-point $\theta_I$. 

For $\tau^{\rm syn}_I$ and $\gamma$ large enough, such that there are a reasonably high rate of spikes and $s_i(t)$ is effectively averaging the spikes over time, so we replace the spikes by the rate $\phi_I(V)$. 
This step is heuristic: the time-average is not equivalent to a trial average because of the non-equilibrium behavior of the network, but for intermediate $\tau^{\rm syn}_I$ we expect the difference to not be too large.

Then, for $\tau_I \rightarrow 0$, the membrane potential of the spikes essentially follows the input to the cell:
\begin{align*}
V_i(t) &\approx \mathcal E_I + \sum_{j \in E} J_{ij} \dot{n}_j(t) + \sum_{j \in I} J_{ij} \phi_I(V_j(t))\\
&= \mathcal E_I + \sum_{j \in E} J_{ij} \dot{n}_j(t) + \gamma \sum_{j \in I} J_{ij}(V_j - \theta_I ),
\end{align*}
where we have assumed that $\mathcal E_I $ keeps $V_i(t)$ above $\theta_I$ so that the argument of the nonlinearity is positive.

Next, we assume all-to-all connectivity between the populations, and within the inhibitory population \footnote{We neglect the self-coupling term of the inhibitory neurons, which in the reduced model would just shift the denominator of the population average $\psi_I$.}:
\begin{align*}
V_i(t) &= \mathcal E_I + \frac{J_{IE}}{N_E}\sum_{j \in E}  \dot{n}_j(t) + \gamma \frac{J_{II}}{N_I} \sum_{j \in I} (V_j - \theta_I );
\end{align*}
we can solve for the population mean $\psi_I \equiv N_I^{-1} \sum_{i \in i} V_i(t)$ by summing over $i \in I$, giving
\begin{align*}
\psi_I &= \frac{\mathcal E_I-\gamma J_{II} \theta_I}{1 - \gamma J_{II}}+ \frac{J_{IE}}{N_E(1-\gamma J_{II})}\sum_{j \in E}  \dot{n}_j(t).
\end{align*}

If the excitatory neurons have the same type of synaptic input $\sum_{j \in I} J_{ij} s_j(t) \approx \frac{J_{EI}}{N_I} \sum_{j \in I} \gamma(V_j(t) - \theta_I),$ then the excitatory dynamics becomes
\begin{align}
\tau \frac{dV_i(t)}{dt} &\approx -V_i  + \left( \mathcal E - \gamma J_{EI}\theta_I + \gamma J_{EI} \frac{\mathcal E_I-\gamma J_{II} \theta_I}{1 - \gamma J_{II}} \right) \nonumber\\
& ~~~~ + \sum_{j \in E} \left(J_{ij} + \frac{1}{N_E}\frac{\gamma J_{EI}J_{EI}}{1 - \gamma J_{II}} \right) \dot{n}_j(t)
\end{align} 
This is the random regular network with all-to-all inhibitory connections that we consider in Sec.~\ref{sec:randreg}.
We may set, $\gamma J_{II} = -1$, $\mathcal E_I = -\gamma J_{II}\theta_I = +\theta_I$, and $\gamma J_{EI}J_{IE}/(1-\gamma J_{II}) = \gamma J_{EI}J_{IE}/2 = -J(k-2\sqrt{k-1})$ to obtain exactly the model considered.
In the simulations shown in Fig.~\ref{fig:widomcollapse_randreg_sims}I-L, we choose $J_{IE} = J(k-2\sqrt{k-1})$ and $J_{EI} = J_{II} = -0.05 J_{IE}$, with $\tau_I = 0.1$ and $\tau^{\rm syn}_I = 2$, and $\theta_I = -1.0$.
We chose the inhibitory connections to not be equal to the excitatory connections so that the excitatory population drives activity in the inhibitory population.

\section{Renormalization group analysis of the stochastic spiking network model}
\label{sec:NPRG}

In this section we give an overview of the renormalization group method we developed to analyze the spiking network model and obtain the renormalized scaling theory.
We will first give a conceptual overview of the key ideas of the renormalization group, before describing technical details of the method.

\subsection{Renormalization in concept}

Conceptually, the renormalization group (RG) method describes how our mathematical descriptions of physical processes change when we view them at a hierarchy of coarser and coarser scales. 
The prototypical example of a hierarchy of scales in physics is length scale: as we ``zoom out'' from the microscopic scale, processes or structures with fine detail become less resolved, blurring into an overall larger picture.
Depending on the state of the system, as we zoom out a system with a mix of microscopic disordered and ordered regions may look increasingly ordered or increasingly disordered on coarser scales.
At a critical point it may be hard to tell the difference: at each zoom level the system looks statistically similar, owing to the (incomplete) scale invariance caused by processes of all orders being strongly coupled.

\begin{figure*}
 \centering
\includegraphics[scale=0.72]{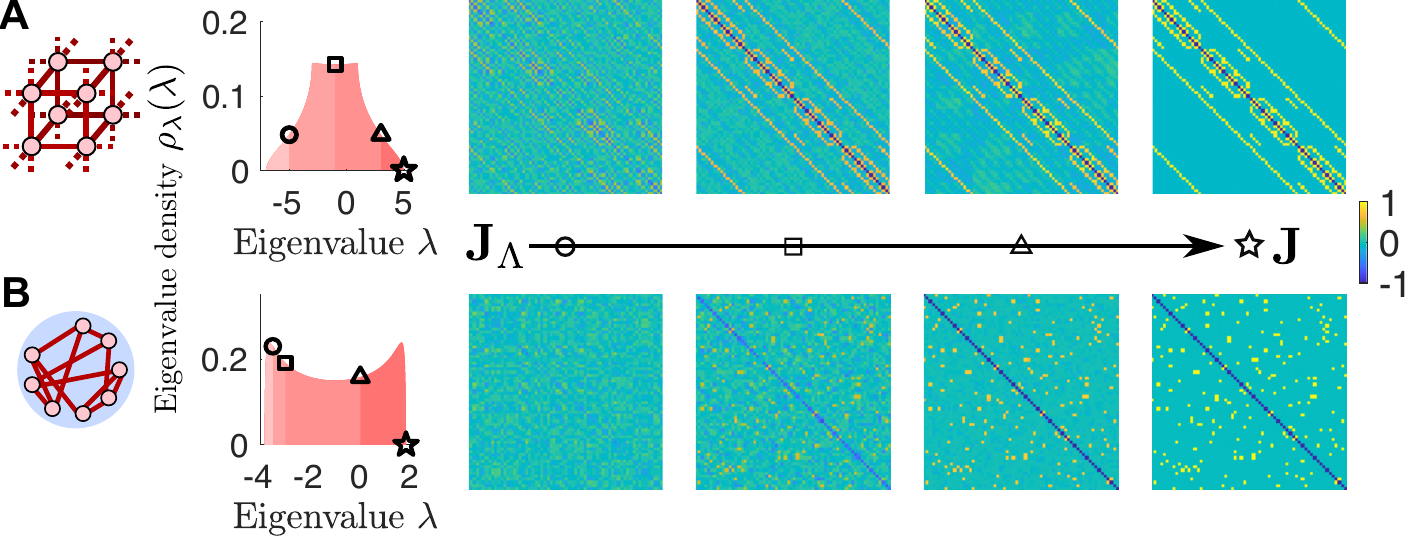}
  \caption{\textbf{Regulation of the synaptic weight matrix by thresholding eigenvalues}. Our renormalization group flow is implemented by thresholding the eigenvalues of the synaptic weight matrices, e.g., for \textbf{A)} 
a 3-dimensional lattice, and \textbf{B)} random regular networks in which the number of connections each neuron makes is fixed but the pairs of neurons connected are otherwise random. The thresholding procedure sets all eigenvalues greater than a threshold $\Lambda$ equal to zero. The full network is recovered as $\Lambda \rightarrow \Lambda_{\rm max}$. Note that, as in the case of the random regular network, it is possible that some eigenvalues are isolated away from the continuous spectrum. These eigenvalues, however, can be moved by appropriate modifications of the network connections; see Sec.~\ref{sec:randreg}.}
  \label{fig:eigenvalueregulation}
% \vspace{-.1in}
\end{figure*}

A challenge for applying the RG to neural circuitry is that is it not clear what the appropriate hierarchy of scales is. 
Coarse graining over spatial scales may not be informative, as connections between neurons may be long range and so such a coarse graining may erroneously blur together neurons whose activities are not strongly coupled.
Data driven approaches, like the phenomenological renormalization group developed by Ref.~\cite{meshulam2019coarse}, have proposed using the eigenvalues of the spike-spike correlations---the ``principal components---as the coarse-graining space, and have shown that the method appears to yield good scaling collapses of hippocampal data.
However, the difficulty with using principal components in a theoretical analysis is that it requires we solve the model first so that we can compute the principal components.
Instead, in this work we show that one fruitful choice of hierarchy of scales is the eigenvalue distribution of the synaptic weight matrix $J_{ij}$. 
If the synaptic connections between neurons are organized in a lattice, then this choice effectively reduces to coarse-graining over ``energy shells,'' which is related to the classic momentum-shell renormalization scheme common in physics.
We illustrate the effect of coarse-graining the eigenspace of the synaptic connections in Fig.~\ref{fig:eigenvalueregulation}. 
In the RG scheme we use in this work, which is based on the ``non-perturbative renormalization group'' (NPRG) method, eigenmodes up to a given eigenvalue threshold $\Lambda$ are progressively incorporated into estimates of the network statistics.
When $\Lambda < \Lambda_{\rm min}$, the smallest eigenvalue of the synaptic weight matrix, the network is decouples into independent Poisson neurons, whose statistics can be evaluated exactly.
As $\Lambda \rightarrow \Lambda_{\rm max}$, all modes are incorporated. 
The RG method provides a means of relating the network statistics at a particular value of $\Lambda$ to the statistics at $\Lambda-d\Lambda$, providing an interpolation from the independent network to the network of interest.

Developing a renormalization group scheme to study spiking network models is non-trivial, and to date has not been done for the leaky integrate-and-fire model considered here, only for network models of units with ``active'' and ``quiescent'' states that can be loosely interpreted as ``spiking'' and ``non-spiking'' \cite{buice2007field,bressloff2010metastable}, or for networks characterized by coarse-grained firing rates, rather than spiking activity \cite{stapmanns2020self,tiberi2022gell}.

The RG method we use in this work is based on the non-perturbative renormalization group (NPRG) method, which has been successfully used to study many problems in condensed matter physics \cite{CanetPRL2004,canet2005nonperturbative,MachadoPRE2010,CanetJPhysA2011,RanconPRB2011,WinklerPRE2013,HomrighausenPRE2013,KlossPRE2014,balog2015activated,canet2015fully,JakubczykPRE2016,DuclutPRE2017}. 
For pedagogical introductions, see \cite{delamotte2012introduction} for equilibrium systems, \cite{CanetPRL2004,canet2006reaction,CanetJPhysA2007,CanetJPhysA2011} for non-equilibrium systems, and \cite{dupuis2021nonperturbative} for a broad overview. 
However, because these methods have been developed for lattices or continuous media in which the fluctuations are driven by Gaussian noise, they cannot be straightforwardly applied to spiking network models. 

In order to implement the RG scheme in practice, we will formulate the stochastic system of equations for the network as a statistical field theory, to which the methods of the non-perturbative renormalization group can be adapted and applied.

\subsection{Renormalization in practice}
\label{sec:RGpractice}

\subsubsection{Field theoretic formulation of the spiking network}
\label{sec:fieldtheorydefn}

We convert Eqs.~(\ref{eqn:LIF})-(\ref{eqn:poiss}) into a field theory using the Martin-Siggia-Rose-Janssen-De Dominicis (MSRJD) path integral formalism for stochastic differential equations \cite{OckerPloscb2017,KordovanArxiv2020,brinkman2018predicting,ocker2023republished}. The probability of the joint membrane and spiking dynamics can be represented as a functional integral $P[V,\dot{n}] = \int \mathcal D[\tilde{V},\tilde{n}]~\exp(-S[\tilde{V},V,\tilde{n},\dot{n}])$, defining the action \cite{brinkman2018predicting,ocker2023republished}
\begin{widetext}
\begin{align}
S[\tilde{V},V,\tilde{n},\dot{n}] &= \sum_{i=1}^N \int_{-\infty}^\infty dt~\left\{\tilde{V}_i(t) \left( \tau \frac{dV_i(t)}{dt} + V_i(t) - \mathcal E - \sum_{j=1}^N J_{ij}\dot{n}_j(t)\right) + \tilde{n}_i(t)\dot{n}_i(t) - \left( e^{\tilde{n}_i(t)} -1 \right)  \phi\left(V_i(t)\right)\right\};
\label{eqn:action}
\end{align}
\end{widetext}
$\tilde{V}$ and $\tilde{n}$ and the auxiliary ``response fields'' that arise in the construction of the path integral. 
The term $\left( e^{\tilde{n}_i(t)} -1 \right) \phi\left(V_i(t)\right)$ arises from choosing the conditional spike probabilities to be Poisson or Bernoulli. 
We do not explicitly write the terms corresponding to initial conditions, as these can be implemented through the source terms to be introduced.
To lighten notation going forward, we will use the shorthands $a \cdot b = \sum_{i,\alpha} \int dt~ a^\alpha_i(t) b^\beta_i(t)$ and $a \cdot M \cdot b = \sum_{ij,\alpha\beta} \int dt dt'~ a^\alpha_i(t) M^{\alpha\beta}_{ij}(t-t') b^\beta_j(t')$, where $i,j$ run over neuron indices, $\alpha,\beta$ index the different fields $\{\tilde{V},V,\tilde{n},\dot{n}\}$ (or their corresponding sources, to be introduced), and $t,t' \in \mathbb{R}$ are times.

The mean-field theory of the model, Eqs.~(\ref{eqn:meanfielddynamics}), formally corresponds to a saddle-point approximation of the probability $P[V,\dot{n}]$, taking variational derivatives of the action with respect to each of the four types of fields, with the steady-state condition imposing $\tilde{V} = \tilde{n} = 0$.
This field theory was first developed for the spiking dynamics (marginalized over $V,\tilde{V}$) by \cite{OckerPloscb2017}, who also developed diagrammatic rules for calculating the perturbative corrections to the mean-field approximation.
This perturbative formalism is useful for improving predictions of network statistics in parameter regimes far from phase transitions, but accurately predicting statistics in the vicinity of a transition requires a renormalization group approach to extend the validity of perturbative approaches to the critical point.
Typical perturbative RG treatments in statistical physics rely on the interactions between units being translation invariant, such that the field theory can be Fourier transformed into momentum-space, allowing for integrating out modes within narrow momentum bands, facilitating perturbative calculation of the RG flow equations.
While many models of neural activity have been formulated using lattices or translation-invariant connections, these are seldom realistic models of neural wiring, and it would be desirable to have an RG formalism that does not rely on translation invariance.
For this purpose we turn to the NPRG.

\subsubsection{The non-perturbative renormalization group extended to the spiking network model}
\label{sec:NPRGspiking}

The key mathematical idea behind the NPRG method is to define a one-parameter family of models that interpolates from a solvable limit of the model to the full theory by means of a differential equation that is amenable to tractable variational approximations, rather than relying on perturbative approximations. 

The variation of the NPRG we adapt starts by modifying the moment generating functional (MGF) $\mathcal Z[\mathcal A]$ or the related cumulant generating functional (CGF) $\mathcal W[\mathcal A] = \ln \mathcal Z[\mathcal A]$,
\begin{align}
    \mathcal Z[\mathcal A] &\equiv \exp\left(\mathcal W[\mathcal A]\right) \label{eqn:MGF} \\
    &= \int \mathcal D[\tilde{V},V,\tilde{n},\dot{n}]~e^{-S[\tilde{V},V,\tilde{n},\dot{n}] + \tilde{V} \cdot h + V \cdot \tilde{h} + \tilde{n} \cdot j + \dot{n} \cdot \tilde{j}}. \nonumber
\end{align}
The MGF and CGF are functionals of ``source fields'' $\mathcal A = \{h, \tilde{h}, j, \tilde{j}\}$. Note that we use the convention of pairing fields with tildes to their partners without tildes, as all fields with tildes may be taken to be purely imaginary. 
Derivatives of the MGF evaluated at zero sources would yield statistical moments and response functions of the joint spike train and membrane potential statistics, while derivatives of the CGF yield cumulants or centered moments and response functions. 

Computing $\mathcal Z[\mathcal A]$ or $\mathcal W[\mathcal A]$ exactly would therefore constitute an exact solution of the stochastic spiking model.
In practice, this is intractable, except in special cases.
One such special case is the limit of no synaptic coupling: $J_{ij} = 0$.
In this case the neural spikes are just independent Poisson processes driven by membrane potentials that sit at the baseline $\mathcal E$.
This motivates the choice of regulating the synaptic weights between neurons by replacing $J_{ij}$ with $J_{ij;\Lambda}$, where $\Lambda$ parametrizes the family of models that interpolates from the network of independent neurons to the actual network we want to study.

Because we need only interpolate between these two endpoints, there are many choices we could make for $\Lambda$.
For the symmetric networks we consider in this work a natural choice is to use $\Lambda$ as a threshold on the eigenvalues of the synaptic weight matrix, defining $J_{ij;\Lambda}$ by setting to $0$ any eigenvalues greater than $\Lambda$.
This choice defines a family of MGFs, $\mathcal Z_\Lambda[\mathcal A]$, indexed by the value of the eigenvalue threshold $\Lambda \in [\Lambda_{\rm min},\Lambda_{\rm max}]$, where $\Lambda_{\rm min}$ and $\Lambda_{\rm max}$ are the smallest and largest eigenvalues of the synaptic weight matrix.
We can relate the MGF at one value of $\Lambda$ to its value at $\Lambda + d\Lambda$ by taking a partial derivative of the definition (\ref{eqn:MGF}). 
The derivative will bring down a factor of $\sum_{ij} \int dt dt'~\tilde{V}_j(t') \partial_\Lambda J_{\Lambda;ij}(t-t')\dot{n}_j(t')$ inside the path integral; the factors of $\tilde{V}_i(t)$ and $\dot{n}_j(t')$ can be replaced with variational derivatives with respect to their conjugate sources, allowing us to pull the differential operator outside of the path integral, giving
\begin{align}
\partial_\Lambda \mathcal Z_\Lambda[\mathcal A] = \sum_{ij} \int dt dt'~\partial_\Lambda J_{\Lambda;ij}\frac{\delta^2 \mathcal Z_\Lambda[\mathcal A]}{\delta h_i(t) \delta \tilde{j}_j(t')}.
\label{eqn:Zfloweqn}
\end{align}
While this looks like a kind of linear differential equation for the functional $\mathcal Z_\Lambda[\mathcal A]$, in practice it is actually more useful to transform this into a flow equation for a related object, the \emph{average effective action} (AEA) $\Gamma[\chi]$.
The regulated AEA is defined as a modified Legendre transform of the CGF $\mathcal W_\Lambda[\mathcal A]$, 
\begin{align}
\Gamma_\Lambda[\chi] &= -\mathcal W_\Lambda[\mathcal A] + \chi \cdot \mathcal A - \frac{1}{2} \chi \cdot [J - J_\Lambda] \cdot \chi,
\label{eqn:GammaLamdefn}
\end{align}
where $\chi = \{\tilde{\psi},\psi,\tilde{\nu},\nu\}$ can be thought of as the expected values of the fields $\{\tilde{V},V,\tilde{n},\dot{n}\}$, respectively,  in the presence of the sources.
The conjugate sets of fields are defined as functions of each other via the relations
\begin{align}
\chi_i(t) = \frac{\delta \mathcal W[\mathcal A]}{\delta \mathcal A_i(t)},&~{\rm or}~\mathcal A_i(t) = \frac{\delta \Gamma[\chi]}{\delta \chi_i(t)}, \label{eqn:chiArelations}
\end{align}
allowing conversion between the CGF and the AEA.
The term $J - J_\Lambda$ in Eq.~(\ref{eqn:GammaLamdefn}) couples only $\tilde{\psi}$ and $\nu$ fields.
By construction, $\Gamma_{\Lambda = \Lambda_{\rm min}} = S[\chi]$ is the mean-field theory of the spiking network model and $\Gamma_{\Lambda = \Lambda_{\rm max}}[\chi] = \Gamma[\chi]$, the true AEA of the model. 
Note that while the mean-field equations (\ref{eqn:meanfielddynamics}) derive from the saddle-points of the action $S$, the equations of motion for the \emph{true} means are saddle-points of the AEA $\Gamma$.

Adapting the derivation in Ref.~\cite{CanetJPhysA2007}, we can show the AEA obeys the celebrated Wetterich flow equation \cite{WetterichPhysLettB1993}, 
\begin{equation}
\partial_\Lambda \Gamma_\Lambda = \frac{1}{2}{\rm Tr}\left[ \partial_\Lambda \mathbf{R}_\Lambda \cdot \left[ \Gamma^{(2)}_\Lambda + \mathbf{R}_\Lambda\right]^{-1}\right],
\label{eqn:Wettericheqn}
\end{equation}
where ${\rm Tr}$ denotes a super-trace over field indices $\chi$, neuron indices, and times. 
$\Gamma^{(2)}_\Lambda$ is a $4N \times 4N$ matrix of second derivatives of $\Gamma_\Lambda$ with respect to pairs of the fields $\chi$, $\mathbf{R}_\Lambda = \mathbf{J} - \mathbf{J}_\Lambda$, and the factor $\left[ \Gamma^{(2)}_\Lambda + \mathbf{R}_\Lambda\right]^{-1}$ is an inverse taken over matrix indices, field indices, and time. 

The Wetterich equation is exact, but being a functional integro-partial differential equation it cannot be solved in practice, and approximations are still necessary. 
Despite the flow equation for (\ref{eqn:Wettericheqn}) appearing much more complicated than (\ref{eqn:Zfloweqn}), the advantage of using $\Gamma_\Lambda[\chi]$ over $\mathcal Z_\Lambda[\mathcal A]$ is that the AEA shares much of its structure with the original action $S$, allowing us to better constrain our non-perturbative approximation. 
The standard approach is to make an \emph{ansatz} for the form of the solution, constrained by symmetries or Ward-Takahashi identities, and employing physical intuition. 
The action of the spiking network model does not readily admit any obvious symmetries, but we can derive a pair of Ward-Takahashi identities that allows us to restrict the form of the AEA. 

The common approach to deriving WT identities is to perturb a field by an infinitesimal amount and demand the resulting linear variation in the action vanishes \cite{canet2015fully}.
However, an alternative approach is available for the spiking model.
In the spiking model we can analytically integrate out either the membrane potential fields or the spiking fields when evaluating the MGF, leaving a path integral over the remaining pair of fields to be performed.
If we integrate out the membrane potential fields and then differentiate the MGF with respect to $\tilde{h}_i(t)$, we obtain the identity
\begin{align*}
\left(\tau \frac{d}{dt}+1\right)\frac{\delta \mathcal Z[\mathcal A]}{\delta \tilde{h}_{i}(t)} &= (h_i(t) + \mathcal E) \mathcal Z[\mathcal A] + \sum_{j}J_{ij} \frac{\delta \mathcal Z[\mathcal A]}{\delta \tilde{j}_{j}(t)}.
\end{align*}
If we integrate out the spike fields and differentiate with respect to the source $\tilde{j}_i(t)$, we obtain the identity
\begin{align*}
\frac{\delta \mathcal Z[\mathcal A]}{\delta j_{i}(t)} &= \tilde{j}_{i}(t)Z[\mathcal A] + \sum_{j} \frac{\delta \mathcal Z[\mathcal A]}{\delta h_{j}(t)}J_{ji}.
\end{align*}
Using the definition $\mathcal Z_\Lambda = \exp(\mathcal W_\Lambda)$ to write these identities in terms of the CGF $\mathcal W_\Lambda$ and then using the relationships (\ref{eqn:chiArelations}) to replace sources with variational derivatives of $\Gamma$ and derivatives of $\mathcal W$ with the expectation fields, we conclude that the AEA must have the form
\begin{widetext}
\begin{equation}
\Gamma[\tilde{\psi},\psi,\tilde{\nu},\nu] = \sum_{i=1}^N \int_{-\infty}^\infty dt~\left\{\tilde{\psi}_i(t) \left(\tau \frac{d\psi_i(t)}{dt} + \psi_i(t) - \mathcal E_i - \sum_{j=1}^N J_{ij} \nu_j(t)\right) + \tilde{\nu}_i(t) \nu_i(t)\right\} + \Upsilon[\tilde{\nu},\psi],
\label{eqn:Gammakappa}
\end{equation}
\end{widetext}
where $J$ is the true synaptic coupling, not the regulated coupling $J_\Lambda$, and the functional $\Upsilon[\tilde{\nu},\psi]$ couples only the spike-response fields $\tilde{\nu}$ and the membrane-potential fields $\psi$. 

Our result for Eq.~(\ref{eqn:Gammakappa}) shows that the membrane dynamics are unrenormalized by stochastic fluctuations---only the interactions between the membrane potential and the spiking statistics are renormalized, and we need only derive the RG flow for the functional $\Upsilon_\Lambda[\tilde{\nu},\psi]$.
To do so, we exploit the fact that the networks we consider in this application have a leading homogeneous mode (i.e., $\sum_{j=1}^N J_{ij} = J\Lambda_{\rm max}$).
We follow previous NPRG work by performing a ``local potential approximation (LPA)'' in which we set the fields to time- and index-independent values $\tilde{\nu}_i(t) = \tilde{\nu}$ and $\psi_i(t) = \psi$.
This reduces $\Upsilon$ to a function we need to solve for, not a functional.
We define the ``local potential'' $U_\Lambda$ by
\begin{equation}
\Upsilon_\Lambda[\tilde{\nu},\psi]\Big|_{\tilde{\nu}_i(t) = \tilde{\nu},~\psi_i(t) = \psi} \equiv -NT U_\Lambda(\tilde{\nu},\psi),
\label{eqn:Upsilonansatz_appendix}
\end{equation}
where square brackets denote functionals and round brackets denote a function.
The proportionality factors $N$ and $T$ are the number of neurons and the duration of the spike train, which become infinite but cancel out of the flow equation.

Using this approximation, we compute the functional derivatives of $\Gamma_\Lambda$ in the matrix $\Gamma^{(2)}$, \emph{and then} evaluate them at homogeneous values $\tilde{\nu}_i(t) \rightarrow \tilde{\nu}$ and $\psi_i(t) \rightarrow \psi$. 
After inserting the homogeneous fields, it is possible to invert $\mathbf{\Gamma}^{(2)}_\Lambda + \mathbf{R}_\Lambda$  in closed form. 
The super-trace ${\rm Tr}$ involves a sum over the four fields, neural indices, and temporal frequency.
For symmetric matrices $\mathbf{J}$ the matrices can be diagonalized, and the cyclic nature of the trace causes the eigenvectors to drop out of the flow equation.
The result in the $N \rightarrow \infty$ limit is 
\begin{widetext}
\begin{align}
\partial_\Lambda U_\Lambda(\tilde{\nu},\psi) &=  \frac{1}{2\tau}\rho_{\lambda}(\Lambda)  \Bigg\{1 -\Lambda U^{(1,1)}_\Lambda(\tilde{\nu},\psi) - \sqrt{\left(1- \Lambda U^{(1,1)}_\Lambda(\tilde{\nu},\psi)\right)^2 - \Lambda^2U^{(0,2)}_\Lambda(\tilde{\nu},\psi)U^{(2,0)}_\Lambda(\tilde{\nu},\psi)} \Bigg\}, \label{eqn:Ufloweqn1loopform}
\end{align}
\end{widetext}
where $\rho_{\lambda}(\lambda)$ is the eigenvalue density of $J_{ij}$ and the initial condition is $U_{\Lambda = \Lambda_{\rm min}}(\tilde{\nu},\psi) = (e^{\tilde{\nu}}-1)\phi(\psi)$. 
Note that in the remainder of this appendix we absorb the synaptic strength $J$ into the definition of the eigenvalues.

In practice, we do not solve Eq.~(\ref{eqn:Ufloweqn1loopform}) directly. 
Instead, we expand $U_\Lambda(\tilde{\nu},\psi) = \sum_{m=1}^\infty \frac{\tilde{\nu}^m}{m!} \Phi_{m,\Lambda}(\psi)$, introducing an infinite set of effective nonlinearities $\Phi_{\Lambda,m}(\psi)$.
We obtain a hierarchy of partial differential equations for these nonlinearties by differentiating Eq.~(\ref{eqn:Ufloweqn1loopform}) $m$ times with with respect to $\tilde{\nu}$ and then setting $\tilde{\nu} = 0$. 
This procedure will yield the hierarchy of flow equations of the form 
\begin{align}
\partial_\Lambda \Phi_{m,\Lambda}(\psi) &= \rho_\lambda(\Lambda)\mathcal F_m(\Phi_{1,\Lambda},\dots,\Phi_{m,\Lambda},\Phi_{m+1,\Lambda}). \label{eqn:Phim}
\end{align}
The functions $\mathcal F_m$ depend on the nonlinearities as well as derivatives of those nonlinearities with respect to $\psi$, which are not denoted explicitly as arguments.
The primary nonlinearity of interest will be the nonlinearity $\Phi(\psi) \equiv \Phi_{1,\Lambda_{\rm max}}(\psi)$, which appears in the renormalized equations for the dynamics of the trial-and-population-averaged means introduced in Sec.~\ref{sec:beyondmft}, Eqs.~(\ref{eqn:truemeandyn})-(\ref{eqn:truenu}).
However, to understand the origin of anomalous critical exponents at the phase transitions in both the \emph{in vitro} and \emph{in vivo} networks, it will be necessary to compute contributions from higher order nonlinearities.

The hierarchy at level $m$ depends only on the next order nonlinearity $\Phi_{m+1,\Lambda}(\psi)$. 
This provides a convenient means of closing the hierarchy and solving a finite system of equations for the nonlinearities.
We close the hierarchy at order $m$ by making the approximation $\Phi_{m+1,\Lambda}(\psi) \approx \Phi_{1,\Lambda}(\psi)$. 
This closure approximation is motivated by the idea that closing the hierarchy at $m=2$ would amount to the approximation $U_\Lambda(\tilde{\nu},\psi) = (e^{\tilde{\nu}}-1) \Phi_{1,\Lambda}(\psi)$; i.e., the nonlinearity is renormalized but not the ``Poisson-ness'' of the spiking fluctuations.
Cutting the hierarchy off at higher orders $m$ amounts to truncating an expansion of $U_\Lambda$ in powers of $e^{\tilde{\nu}}-1$, effectively limiting how non-Poisson the spiking becomes.

\begin{figure*}
 \centering
  \includegraphics[scale=1]{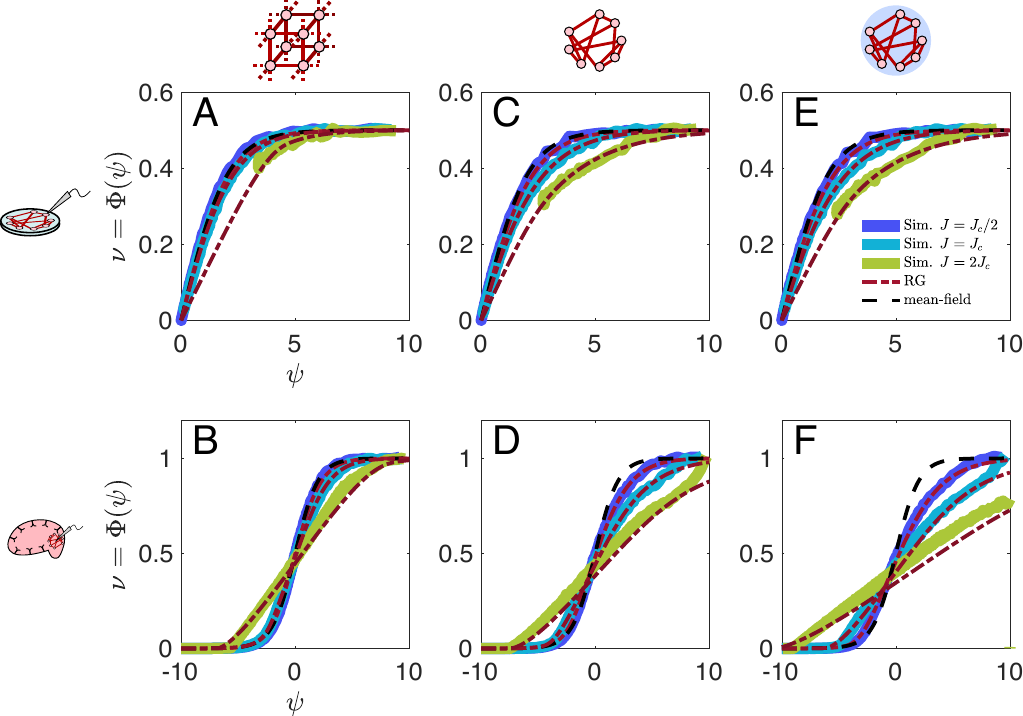}
  \caption{\textbf{Effective nonlinearities on different networks with effective dimension $d = 3$} and synaptic strengths $J = \{J_c/2,J_c,2J_c\}$. \textbf{A-B)} Effective nonlinearities for a cubic lattice of $N = 14^3$ neurons for an absorbing state network (\textbf{A}) with $J_c \approx 0.825$ and a spontaneous network (\textbf{B}) with $J_c \approx 1.165$. \textbf{C-D)} Effective nonlinearities for an excitatory random regular network of $N = 2^{11}$ neurons for an absorbing state network (\textbf{C}) with $J_c \approx 2.17$ and a spontaneous network (\textbf{D}) with $J_c \approx 3.75$. \textbf{E-F)} Effective nonlinearities for an effective excitatory-inhibitory population with sparse excitatory-excitatory connections and dense (all-to-all) connections between other pairs, with $N = 2^{11}$ neurons, for an absorbing state network (\textbf{E}) with $J_c \approx 2.32$ and a spontaneous network (\textbf{F}) with $J_c \approx 6.0$. In all cases the blue data points are simulated data averaged over $100$ trials.  The red curves are the predictions of the hierarchy of nonlinearities (Eqs.~(\ref{eqn:Phim})) with the truncation $\Phi_{5,\Lambda}(\psi) \approx \Phi_{1,\Lambda}(\psi)$. The black dashed lines correspond to the mean-field prediction, $\Phi_{1,\Lambda}(\psi) = \phi(\psi)$. The curves in the supercritical phases are less accurate because the solution becomes non-analytic, which is difficult to capture numerically.}
  \label{fig:networknonlinearities}
\end{figure*}

As a concrete example, the flow equations for $\Phi_{1,\Lambda}(\psi)$  and $\Phi_{2,\Lambda}(\psi)$  work out to be
\begin{widetext}
\begin{align}
    \partial_\Lambda\Phi_{1,\Lambda}(\psi) &= \frac{\rho_\lambda(\Lambda)\Lambda^2}{4\tau} \frac{ \Phi_{2,\Lambda}(\psi) \Phi_{1,\Lambda}''(\psi)}{1 - \Lambda \Phi_{1,\Lambda}'(\psi)} \label{eqn:Phi1flow}\\ 
\partial_\Lambda\Phi_{2,\Lambda}(y) &= \frac{ \rho_\lambda(\Lambda)\Lambda^2}{8\tau} \left[\frac{\Lambda^2 \Phi_{2,\Lambda}(y)^2 \Phi_{1,\Lambda}''(y)^2}{(1 - \Lambda \Phi_{1,\Lambda}'(y))^3}+\frac{4\Lambda \Phi_{2,\Lambda}(y)\Phi_{2,\Lambda}'(y)\Phi_{1,\Lambda}''(y)}{(1 - \Lambda \Phi_{1,\Lambda}'(y))^2} + \frac{4\Phi_{3,\Lambda}(y)\Phi_{1,\Lambda}''(y) + 2 \Phi_{2,\Lambda}(y)\Phi_{2,\Lambda}''(y)}{1 - \Lambda \Phi_{1,\Lambda}'(y)}\right]. \label{eqn:Phi2flow}
\end{align}
\end{widetext}
We remind the reader that in these equations we have absorbed the synaptic strength $J$ into the definition of the eigenvalue $\Lambda$.
The initial initial conditions are $\Phi_{m,\Lambda_{\rm min}}(\psi) = \phi(\psi)$, with boundary conditions $\Phi_{m,\Lambda}(\psi) \overset{\lim |\psi| \rightarrow \infty}{\sim} \phi(\psi)$ for \emph{in vivo} networks and $\Phi_{m,\Lambda}(0) = 0$ for \emph{in vitro} networks, where we take $\psi = 0$ to be the activation threshold. 
The numerical solutions for $\Phi(\psi) = \Phi_{1,\Lambda_{\rm max}}(\psi)$ plotted in Fig.~\ref{fig:networknonlinearities} are obtained by numerically solving Eq.~(\ref{eqn:Phi1flow}) up to fourth order ($m=3,4$ equations not shown).

While we cannot solve the hierarchy of equations in closed form, we show next that we can analytically extract the anomalous scaling behavior when the network is tuned close to a critical point.
This anomalous behavior emerges in the solutions of the hierarchy (\ref{eqn:Phim}) when $1 - \Lambda \Phi_{1,\Lambda}'(\psi)$---the denominator in Eq.~(\ref{eqn:Phi1flow}), which appears in all equations in the hierarchy---vanishes.
This factor is finite when the network is in the subcritical regime and $\Lambda < \Lambda_{\rm max}$, but at a critical value of $J = J_c$ there is a point $\psi = \psi_c$ at which the denominator vanishes at the end of the flow.
The overall solution does not diverge, but becomes non-analytic, depending on non-integer powers of $\psi-\psi_c$.
For $J > J_c$ we expect the denominator to remain finite, though possibly small, for $\Lambda < \Lambda_{\rm max}$, but as $\Lambda \rightarrow \Lambda_{\rm max}^-$ the denominator $1 - \Lambda_{\rm max} \Phi_{1,\Lambda}'(\psi)$ vanishes for all $\psi \in [\psi_-,\psi_+]$ and $\Phi_{1,\Lambda_{\rm max}}(\psi)$ becomes exactly linear on this interval, as seen in Fig.~\ref{fig:networknonlinearities}.
The end-points $\psi_\pm$ represent the possible metastable states of the network in the supercritical phase, and the interval $[\psi_-,\psi_+]$ represents a coexistence region between metastable states. 
This is directly analogous to the development of the non-analyticity in the free-energy of the Ising model in the ordered phase \cite{GoldenfeldBook1992}.

\subsubsection{Universality in the renormalization group flow}
\label{sec:universality}

So far, our RG treatment of the spiking network model has implemented the first step of an RG procedure, coarse-graining. 
To extract the anomalous scaling behavior of the nonlinearities close to the critical point, we need to implement the second RG step, rescaling.

In the NPRG context, the rescaling procedure will amount to identifying an appropriate non-dimensionalization of the hierarchy of flow equations Eq.~(\ref{eqn:Phim}) and searching for fixed point solutions. 
We make the change of variables
\begin{align}
\Phi_{1,\Lambda}(\psi) &= \Pi_s + \Lambda^{-1}_{\rm max}(\psi-\theta_s) \label{eqn:varphi1} \\
& ~~~~~ + e^{-s(d/2+1-\eta^X_s)} \varphi_{1,s}\left((\psi-\theta_s)e^{s(d/2-\eta^X_s)}\right), \nonumber \\
\Phi_{m,\Lambda}(\psi) &= e^{-s(d/2+1- m\eta^X_s)} \nonumber \\
& ~~~~~ \times \varphi_{m,s}\left((\psi-\theta_s)e^{s(d/2-\eta^X_s)}\right), \label{eqn:varphim}
\end{align}
where we define the ``RG-time'' 
\begin{equation}
s \equiv - \ln \left( \frac{\Lambda_{\rm max} - \Lambda}{\Lambda_{\rm max} - \Lambda_{\rm min}} \right) \in [0,\infty);
\label{eqn:RGtime}
\end{equation} 
we define $s$ to be positive, in contrast to the convention in some NPRG works. 
In the first nonlinearity we remove the running baseline firing rate $\Pi_s = \Phi_{1,\Lambda}(\theta_s)$ and the critical slope $\Lambda_{\rm max}^{-1}(\psi-\theta_s)$, as these would become infinite offsets under our rescaling procedure. 
In the \emph{in vitro} networks $\Pi_s = 0$ and $\theta_s = 0$, as stochastic fluctuations do not generate activity-independent spiking, nor do they shift the activation threshold of the effective nonlinearity. 

This change of variables introduces the ``dimensionless'' nonlinearities $\varphi_{m,s}(z)$, which represent centered and rescaled versions of the nonlinearities $\Phi_{m,\Lambda}(\psi)$, where $z \equiv (\psi-\theta_s)e^{s(d/2-\eta^X_s)}$ is the dimensionless membrane potential centered on the running set-point $\theta_s = {\rm argmax}~\Phi_{1,\Lambda}'(\psi)$.
The factor $e^{s(d/2-\eta^X_s)}$ is a ``running'' scale-factor that zooms in on this point, where $d$ is the spectral dimension of the network (Eq.~(\ref{eqn:spectraldimdefn})).
Similarly, the factors $e^{-s(d/2+1-m \eta^X_s)}$ scale down the amplitudes of the nonlinearities $\Phi_{m,\Lambda}(\psi)$; the factor $e^{s m\eta^X_s}$ originates from a rescaling of the spike response field $\tilde{\nu} = \tilde{z} e^{-s\eta^X_s}$, which defines the running exponent $\eta^X_s$, to be determined shortly.
We omit some dimensional constant factors that depend on the eigenvalue density and initial values of the couplings, as these do not contribute to the critical exponents.

Our choice of scaling factors renders the flow equations for $\varphi_{m,s}(z)$ asymptotically independent of $s$ for $s \gg 1$. e.g.,
\begin{widetext}
\begin{align}
    \partial_s\varphi_{1,s}(z) - \left(\frac{d}{2}+1 - \eta^X_s\right)\varphi_{1,s}(z) + \left[\left(\frac{d}{2}-\eta^X_s\right)z - \zeta_s\right]\varphi_{1,s}'(z)  &= \frac{1}{2} \frac{ \varphi_{2,s}(z) \varphi_{1,s}''(z)}{1 - \varphi_{1,s}'(z)},\label{eqn:varphi1flow}
\end{align}
\end{widetext}
with similar but more complicated equations for higher order $\varphi_{m,s}(z)$.
There is also a flow equation for $\Pi_s$, but while it is driven by the flow of the $\varphi_{m,s}(z)$, it does not couple back into the nonlinearities. 
When searching for fixed points of the flow equation we may thus ignore the flow of $\Pi_s$, though it will converge to a (non-universal) value $\nu_c$ at the end of the flow that determines the mean firing rate at the critical point.

In \emph{in vivo} networks the term $\zeta_s \equiv e^{s(d/2-\eta^X_s)}\partial_s \theta_s$ is necessary to fix $\varphi_{1,s}''(0) = 0$ for all $s$ throughout the flow, which follows from centering our rescaling point around $\theta_s = {\rm argmax}~\Phi_1'(\psi)$.
$\zeta_s$ depends on several derivatives of $\varphi_{m,s}(z)$ at $z = 0$, and we will find that at the critical point $\zeta_\ast = 0$.
In \emph{in vitro} networks $\theta_s = 0$, and hence $\zeta_s = 0$.

Similarly, $\eta^X_s$ is chosen to either impose a relationship between derivatives (as we will do for \emph{in vitro} networks) or fix one of the derivatives of the nonlinearites to $1$ (as we will do for \emph{in vivo} networks), and will depend on derivatives of $\varphi_{m,s}(z)$ at $z = 0$.

Although Eq.~(\ref{eqn:varphi1flow}) is only valid for RG-times $s \rightarrow \infty$, we retain some autonomous time-dependence for the purposes of performing linear stability analyses around fixed points of the flow. 
That is, we will expand the dimensionless nonlinearities around their fixed points, $\varphi_{m,s}(z) = \varphi_{m\ast}(z) + e^{\mu_{\ell}s} v_{m,\ell}(z)$, to obtain a system of equations for the eigenmodes of the RG flow, $v_{m,\ell}(z)$, and their associated eigenvalues $\mu_{\ell}$.
If $\mu_\ell > 0$ the eigenmode is ``relevant,'' and the RG flow is repelled away from the critical point along the directions $v_{m,\ell}(z)$.
If $\mu_\ell < 0$ the eigenmode is ``irrelevant,'' and projections of the dimensionless nonlinearities onto the modes $v_{m,\ell}(z)$ will decay.
However, despite this decay, we will show in Sec.~\ref{sec:criticalnonlin} that the irrelevant eigenmodes are important for understanding the shape of the effective nonlinearity $\Phi_{1,\Lambda}(\psi)$.

We cannot solve Eq.~(\ref{eqn:varphi1flow}) and its higher order siblings exactly, as they are nonlinear partial differential equations, so we use a combination of perturbative and non-perturbative techniques to estimate the fixed points.
Our scheme involves first estimating the fixed points with a low-order non-perturbative approximation, which gives a qualitative picture of the transition and determines the appropriate expansion parameter to use in a perturbative solution of our flow equations for dimensions close to the upper critical dimension, the dimension above which the only stable fixed point of the flow equation is trivial.
This perturbative solution can then be used to seed an iterative root-finding routine to solve for the fixed points of a higher order non-perturbative expansion.
Here we only show the minimal truncation to illustrate the qualitative ideas of the truncation procedure, and quote the results of the higher order truncations.

The non-perturbative method involves expanding $\varphi_{m,s}(z)$ in a series around $z=0$ and truncating at a finite order,
\begin{align}
    \varphi_{m,s}(z) &= \sum_{n = 1}^\infty \frac{g_{mn,s}}{n!} z^n,
    \label{eqn:gseries}
\end{align}
where we have defined the dimensionless running couplings $\varphi^{(n)}_{m,s}(0) \equiv g_{mn,s}$. 
Truncating this series does not reflect an assumption that the variable $z$ is small, but constitutes a further variational projection onto a reduced solution subspace, similar to how RG analyses of the Ising model often track only the flow of two couplings, despite coarse graining generating couplings of all orders. 
After choosing a finite number of couplings to consider, we obtain a system of differential equations by differentiating the flow equations for $\varphi_{m,s}(z)$ with respect to the appropriate powers of $z$ and evaluating at $z = 0$.

For a low order truncation in just a few of the $g_{mn,s}$ terms we can analytically solve for the fixed points and determine their dependence on the difference between the upper critical dimension $d_c$ and the spectral dimension $d$.
A linear stability analysis around these fixed points also gives an estimate of the critical exponent $\nu_\ast$.

Our analyses are slightly different for \emph{in vitro} and \emph{in vivo} networks, owing to the choice of $\eta^X_s$, which we consider separately below.

\subsubsection{\emph{In vitro} networks}
\label{subsec:absorbingstatenetworks}

Non-equilibrium models with absorbing states, such as the inactive state of our \emph{in vitro} model, often fall into the directed percolation universality class \cite{janssen2005field}, with exceptions when there are additional symmetries satisfied by the microscopic action \cite{tarpin2017nonperturbative}.
The primary symmetry of the directed percolation  universality class is the ``rapidity symmetry.'' 
Translated into the spiking network model, rapidity symmetry would correspond to an invariance of the AEA under the transformation $\tilde{\nu}_i(t) \leftrightarrow -c \psi_i(t)$, where $c$ is a constant chosen so that the terms $\tilde{\nu}_i(t) \psi_i(t)^2$ and $\tilde{\nu}_i(t)^2 \psi_i(t)$ transform into each other, including their coefficients. 
The spiking network does not obey this symmetry; however, most models in the DP universality class do not exhibit rapidity symmetry exactly: it is instead an emergent symmetry that is satisfied after discarding irrelevant terms in an action tuned to the critical point \cite{henkel2008universality}. 
 
To demonstrate that the spiking network model belong to the directed percolation universality class, we choose the running exponent $\eta^X_s$ to impose the rapidity symmetry relationship on the couplings $g_{21,s}$ and $g_{12,s}$.
Note that because all couplings $g_{mn,s}$ with $n = 0$ are initially zero, they remain so throughout the RG flow, so $g_{21,s}$ and $g_{12,s}$ are the lowest order couplings we may use to fix $\eta^X_s$.
Rapidity symmetry renders $g_{21,s} = - g_{12,s}$ for all $s$, a hallmark of the Reggeon field theory action that describes the directed percolation universality class \cite{janssen2005field,canet2006reaction}. 
We can then show that $\varphi_{m\ast}(z) = 0,~\eta^X_\ast = d/4$ is a trivial fixed point that loses stability below the upper critical dimension $d_c = 4$. 
Below the upper critical dimension $g_{12,s}$ or $g_{21,s}$ can flow to non-zero fixed-point values $g_{12}^\ast = -g_{21}^\ast$ for a fine-tuned value of $J$.
We assume that $\phi'(0^+) > 0$ and $\phi''(0^+) < 0$, which determines that $g_{12,s} < 0$ and hence $g_{21,s} > 0$ \footnote{Although we can choose an initial nonlinearity for which $\phi''(0)$ vanishes, this property is not preserved by the flow of the effective nonlinearity $\Phi_{1,\Lambda}(\psi)$, and in general $\Phi_{1,\Lambda \neq \Lambda_{\rm min}}''(0) \neq 0$. Because the dimensionless flow equation is only asymptotic for $\Lambda \gtrsim \Lambda_{\rm max}$, we may use $\Phi_{1,\Lambda \gtrsim \Lambda_{\rm min}}(\psi)$ in place of $\phi(\psi)$ define this non-dimensionalization instead.}. 
We will focus on $g_{21,s}$ in our presentation of the RG flow in the space of the couplings $g_{11,s}$ and $g_{21,s}$.

The running exponent $\eta^X_s$ can be defined by differentiating the flow equations to derive equations for $g_{12,s}$ and $-g_{21,s}$ and equating them. 
This reveals that
\begin{align}
    \eta^X_s &= \frac{d}{4} + \frac{1}{2} \frac{ g_{13,s} - g_{31,s} }{1 - g_{11,s}}.
    \label{eqn:ASeta}
\end{align}
In general, rapidity symmetry requires $g_{mn}^\ast = (-1)^{m+n} g_{nm}^\ast$ at the fixed point \cite{canet2006reaction}. 
Because the bare action $S[\tilde{V},V,\tilde{n},\dot{n}]$ is not invariant under rapidity symmetry, this relation is not obeyed by $g_{mn,s}$ and $g_{nm,s}$ at finite $s$, but is attained in the $s \rightarrow \infty$ limit.
In this limit, $\eta^X_\ast = d/4$ for all $d$ and the anomalous exponent $\eta_\ast \propto \eta^X_\ast - d/4$ is always $0$ within our approximation.
Note that because rapidity symmetry imposes a relationship between $g_{mn}^\ast$ and $g_{nm}^\ast$, to properly capture such a fixed point we must include both terms in any truncation we make; i.e., if we truncate at order $z^n$, we must include these terms from all nonlinearities up to $\varphi_{m=n,s}(z)$.

\begin{figure*}
 \centering
\includegraphics[scale = 1]{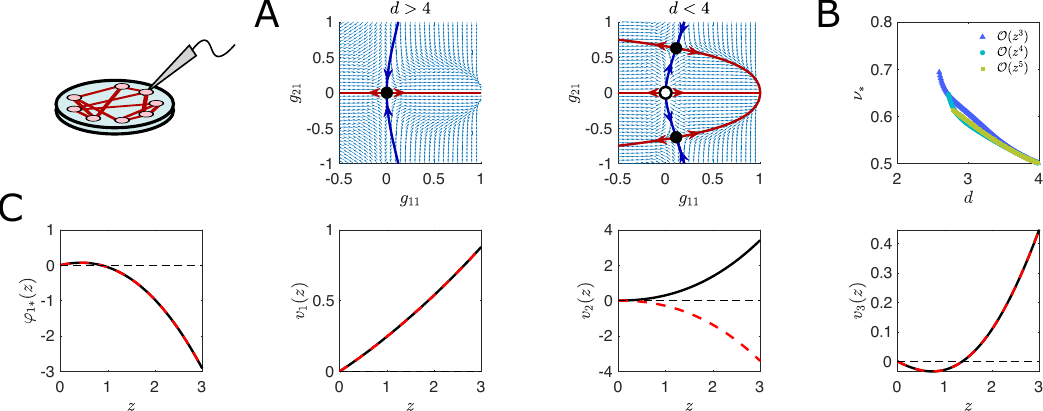}
  \caption{\textbf{A. Renormalization group flow of the absorbing state network model} in the space of the couplings $g_{11}$ and $g_{21}$, for effective dimensions $d > 4$ and $d < 4$, where $4$ is the upper critical dimension. In $d > 4$ only a trivial fixed point exists, while in $d < 4$ two equivalent fixed points exist, with $g_{21}^\ast > 0$ selected by the initial conditions of the network model. The stable and unstable manifolds (solid lines) are colored according to the critical points (saddle nodes), with blue indicating the stable manifold and red indicating unstable manifolds. \textbf{B. Correlation length exponent for the absorbing state network} as a function of the effective dimension $d$. Obtained using the third (dark blue triangles), fourth (light blue circles), and fifth order (green squares) truncations. The non-perturbative analysis appears to break down at $d \lesssim 2.78$ within the local potential approximation. \textbf{C. Critical nonlinearity and eigenmodes of the directed percolation critical point} in $d = 3$, calculated up to order $m,n=5$. Black solid lines correspond to the effective nonlinearity $\varphi_1^\ast(z) = \sum_{n=1}^5 \frac{g_{1n}^\ast}{n!} z^n$ and its eigenmodes, while red dashed curves correspond to $-\sum_{n=1}^5 \frac{g_{n,1}^\ast}{n!}(-z)^n$. If rapidity symmetry is obeyed, the black and red curves will match. We see that rapidity symmetry is obeyed by the critical nonlinearity and several eigenmodes, though some eigenmodes (here, $v_2(z)$) possess a different symmetry, $g_{mn}^\ast = (-1)^{m+n+1}g_{nm}^\ast$. The eigenvalues of the modes are $\mu_1 = (2\nu_\ast)^{-1} \approx 0.85, \mu_2 \approx 0, \mu_3 \approx -0.28$.}
  \label{fig:DPbigfig}
\end{figure*}

We can understand the qualitative features of the RG flow of the \emph{in vitro} networks by considering only the couplings $g_{11,s}$ and $g_{21,s}$.
The flow equations for these two couplings are
\begin{align}
\partial_s g_{11,s} &= g_{11,s} - \frac{1}{2} \frac{g_{21,s}^2}{1-g_{11,s}} \label{eqn:ASg11}\\
\partial_s g_{21,s} &= \frac{4-d}{4}g_{21,s} -\frac{g_{21,s}^3}{(1-g_{11,s})^2} \label{eqn:ASg21}
\end{align}
The RG flow of Eqs.~(\ref{eqn:ASg11})-(\ref{eqn:ASg21}) in the $g_{11}{-}g_{21}$ phase plane above and below the upper critical dimension $d = 4$ is shown in Fig.~\ref{fig:DPbigfig}. 
In $d > 4$ only the trivial fixed point $(g_{11}^\ast,g_{21}^\ast) = (0,0)$ exists, while in $d < 4$ we find the fixed point solution
\begin{align*}
g_{11}^\ast = \frac{4-d}{12-d},~g_{21}^\ast = \frac{4 \sqrt{4-d}}{\sqrt{d^2-24 d+144}},
\end{align*}
with $g_{12}^\ast = -g_{21}^\ast$. 

By performing a linear stability analysis around the fixed points we can estimate the correlation length exponent $\nu_\ast$ from the largest eigenvalue of the stability matrix, $\mu$: $\nu_\ast = (2\mu)^{-1}$. The factor of $1/2$ is included so that the value of $\nu_\ast$ matches the numerical values obtained in prior work in translation invariant systems.
When $d > 4$ the trivial fixed point has one negative and one positive eigenvalue, signaling the fact we must tune only one parameter to arrive at this fixed point. 
The positive eigenvalue is $\mu = 1$, giving $\nu_\ast = 1/2$, as expected. 
In $d < 4$ the trivial fixed point becomes wholly unstable as it splits into the pair of non-trivial fixed points shown in Fig.~\ref{fig:DPbigfig}A, which each have a stable and unstable direction, and the eigenvalue of the flow along the unstable manifold gives the correlation length exponent, which we give for $d \rightarrow 4^-$ is
$\nu_\ast \approx 1/2+ (4-d)/16 - 7/128 (4-d)^2 + \dots.$
The expansion of $\nu_\ast$ near $d = 4^-$ matches the one-loop perturbative estimate of $\nu_\ast$ for the Reggeon field theory \cite{janssen2005field}.

Although within this minimal truncation we obtain an expression for $\nu_\ast$ valid for all $d < 4$, the expression for $\nu_\ast$ begins decreasing non-monotonically as $d$ is lowered below $\sim 3.2$. 
This non-monotonic behavior is an artifact of the truncation, as we confirm by increasing the truncation up to order $z^5$, beyond which the calculations become computationally expensive. 
We indeed still find the non-trivial fixed point with rapidity symmetry, for which we estimate the critical exponent $\nu_\ast$ by a linear stability analysis in the higher dimension phase space.
We confirm that the fixed point with rapidity symmetry is not unstable to some other fixed point lacking that symmetry: the DP fixed point has only a single relevant direction for $2.78 \lesssim d < 4$. 
For $d \rightarrow 2.78^+$ the estimates for $\nu_\ast$ diverge for our $z^3$ and $z^4$ truncations, and numerically generates erroneous complex eigenvalues at lower $d$ in our  $z^5$ truncation, as shown in Fig.~\ref{fig:DPbigfig}B. 
In Sec.~\ref{sec:applications} we showed that the critical exponents of the directed percolation universality class in $d=2$ successfully yield a scaling collapse, indicating that the critical phenomena are still controlled by the DP fixed point in $d < 2.78$. 
It is well-known in the NPRG literature that the LPA tends to break down far below the upper critical dimension, when the value of the anomalous exponent grows larger, so this may be an indication of the breakdown of the LPA for the \emph{in vitro} spiking model.

In Fig.~\ref{fig:DPbigfig}C we plot the critical nonlinearity in $d=3$, estimated from a truncation up to $\mathcal O(z^5)$, along with the corresponding components of the eigenmodes of the linear stability analysis around this fixed point.
We plot these against the function $-\tilde{\varphi}_{1\ast}(-z)$, where $\tilde{\varphi}_{1\ast}(z) = \sum_{m=1}^\infty \frac{g_{m,1}^\ast}{m!} z^m$.
If rapidity symmetry is obeyed by the fixed point, we expect $\varphi_1^\ast(z) = -\tilde{\varphi}_{1\ast}(-z)$, and similarly for the eigenmodes.
We see that one of the eigenmodes does not obey rapidity symmetry, which is to be expected because the bare action $S[\tilde{V},V,\tilde{n},\dot{n}]$ does not obey this symmetry.
This particular eigenmode is \emph{marginal}, having an eigenvalue numerically on the order of $10^{-14}$ (which appears to hold in any dimension $d < 4$), though many irrelevant eigenmodes with $\mu < 0$ also lack rapidity symmetry (not shown).
Empirically, the modes lacking rapidity symmetry appear to obey a different symmetry between coefficients, $v(z) = -\tilde{v}(-z)$.
While it is possible that another critical point exists possessing this symmetry, it appears to be unstable with respect to the DP fixed point, at least in dimensions $2 \leq d \leq 4$. 

\subsubsection{\emph{In vivo} networks}
\label{subsec:spontaneousnetworks}

We now consider the case of spontaneously active networks. 
The fact that there is a membrane potential-independent component of the fluctuations in the spontaneous networks suggests we should choose the running exponent $\eta^X_s$ to keep the coupling $g_{20,s} = 1$ for all $s$.
This condition is tantamount to choosing the Gaussian part of the action to be invariant under the RG procedure. 
To enforce this restriction we find that $\eta^X_s$ must be set equal to 
\begin{align}
     \eta^X_s &= \frac{d+2}{4} - \frac{\zeta_s g_{21,s}}{2} + \frac{1}{4} \frac{g_{22,s}}{1-g_{11,s}},
     \label{eqn:etaXgauss}
\end{align}  
where the condition that $g_{12,s} = 0$ for all $s$ gives that $\zeta_s$ must be equal to
\begin{align}
\zeta_s &= - \frac{1}{2} \frac{g_{14,s} - 2g_{13,s} g_{21,s}}{(1-g_{11,s})g_{13,s}}.
\label{eqn:zetaguass}
\end{align}
The trivial fixed point solution is $\varphi_{m\neq2\ast}(z) = 0, \varphi_{2\ast}(z) = 1,~\eta^X_\ast = \frac{d+2}{4},~\zeta_\ast = 0$.
Note that because of the factor $g_{14,s}/g_{13,s}$, we must carefully show that  $g_{14,s} \rightarrow 0$ faster than $g_{13,s} \rightarrow 0$.

A linear stability analysis of this fixed point shows that the coupling $g_{11,s}$ is unstable at any $d$, and its initial condition must be fine-tuned (by tuning $J$) to bring the network to the critical point.
As $d$ is lowered below $4$ the coupling $g_{13,s}$ becomes relevant, and the celebrated Wilson-Fisher (WF) fixed point emerges from the trivial fixed point. 
The WF fixed point controls the critical properties of the Ising model universality class, and hence \emph{in vivo} networks are in this universality class as well.
The WF fixed point has an inversion symmetry $\mathbb{Z}_2$, which translates into evenness or oddness of the effective nonlinearities, $\varphi_{m\ast}(-z) = (-1)^m \varphi_{m\ast}(z)$.
If this were a symmetry of the bare action $S[\tilde{V},V,\tilde{n},\dot{n}]$, it would correspond to an invariance of the action under the transformation $(\tilde{n},V-\theta) \leftrightarrow (-\tilde{n},-(V-\theta))$.
However, the action (\ref{eqn:action}) does not possess this symmetry: the size of spiking fluctuations at low firing rates are different from those at high firing rates. 
Because symmetry breaking terms are normally relevant, and will drive RG flows away from the critical point, we might worry that the WF fixed point cannot be seen in the spiking network model.
However, our linear stability analysis of the WF fixed point confirms that the WF fixed point does not lose stability when symmetry breaking terms are present.
The $\mathbb{Z}_2$ symmetry is simply an emergent symmetry that holds close to the critical point \footnote{Had we not been careful to define $z$ in terms of the difference between the membrane potential and its running set-point $\theta_s$ then our stability analysis would have found a WF fixed point that was unstable toward a complex-valued ``spinodal'' critical point that may describe critical metastable transitions..}.

We validate our above claims by making a minimal truncation of $\varphi_{1,s}(z) = g_{11,s}z  + g_{13,s}z^3/3! + g_{14,s}z^4/4!$, $\varphi_{2,s}(z) = 1$, for which $\eta^X_s = (d+2)/4$ and $\zeta_s = g_{14,s}/(1-g_{11,s})/g_{13,s}$. 
We include the $g_{14,s}$ term to show that the $\mathbb{Z}_2$-symmetric RG fixed point is not unstable to this mode, despite breaking the symmetry.
We do not include higher order terms in $\varphi_{2,s}(z)$, or any components of $\varphi_{m\geq 3,s}(z)$, as it can be shown that these terms are irrelevant in $d > 2$ and may be neglected in this analysis.

\begin{figure*}
 \centering
      \includegraphics[scale=1]{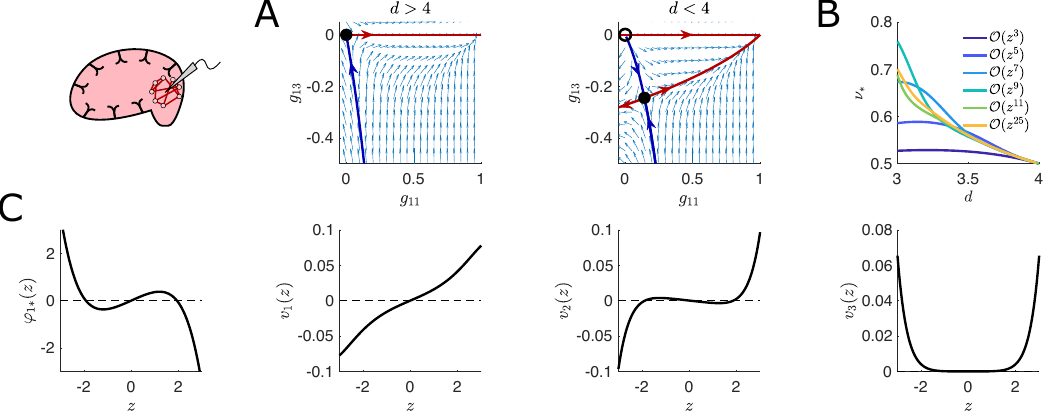}
  \caption{\textbf{A. Renormalization group flow of the spontaneous network model} in the allowed range of the couplings $g_{11}$ and $g_{13}$, for effective dimensions $d > 4$ and $d < 4$, where $4$ is the upper critical dimension. In $d > 4$ only a trivial fixed point exists, while in $d < 4$ the non-trivial Wilson-Fisher fixed point enters the valid range parameter range and has exchanged stability with the trivial fixed point. The stable and unstable manifolds (solid lines) are colored according to the critical points (saddle nodes), with blue indicating the stable manifold and red indicating unstable manifolds. \textbf{B. Correlation length exponent ($\nu^\ast$) estimates for the Wilson-Fisher fixed point} as a function of the effective dimension $d$, obtained using several truncations of the dimensionless nonlinearity $\varphi_1^\ast(z)$ up to order $z^{25}$. \textbf{C. Critical nonlinearity and eigenmodes of the Ising model critical point} in $d = 3$, calculated up to order $m=1,n=11$  The eigenvalues of the modes are $\mu_1 = (2\nu_\ast)^{-1} = 0.73, \mu_2 = -0.18, \mu_3 = -0.81$. At the Ising model fixed point the critical nonlinearity is an odd function $\varphi_1^\ast(-z) = -\varphi_1^\ast(z)$. This symmetry is also obeyed by several eigenmodes, while other eigenmodes are even. Despite breaking the $\mathbb{Z}_2$ symmetry, the even eigenmodes are irrelevant at the fixed point.}
  \label{fig:IMbigfig}
\end{figure*}

The system of equations for the three couplings, plugging in the expressions for $\eta^X_s$ and $\zeta_s$, is
\begin{align}
\partial_s g_{11,s} &= g_{11,s} + \frac{1}{2} \frac{g_{13,s}}{1-g_{11,s}},\\
\partial_s g_{13,s} &= \frac{4-d}{4}g_{13,s} + \frac{3}{2} \frac{g_{13,s}^2}{(1-g_{11,s})^2}\nonumber\\
   & ~~~~~~~~ - \frac{1}{2} \frac{g_{14,s}^2}{(1-g_{11,s})g_{13,s}}\\
\partial_s g_{14,s} &= \frac{10-3d}{4}g_{14,s}+\frac{5g_{14,s}g_{13,s}}{(1-g_{11,s})^2}.
\end{align}
The non-trivial fixed point solution is
\begin{align*}
g_{11}^\ast &= \frac{4-d}{10-d},\\
g_{13}^\ast &= -12\frac{4-d}{(10-d)^2},\\
g_{14}^\ast &= 0.
\end{align*}
We plot the RG flow in the $g_{11}{-}g_{13}$ plane in Fig.~\ref{fig:IMbigfig}A.
The linear stability analysis around this fixed point yields $3$ eigenvalues, of which one is positive and two are negative when $d < 4$.
The eigenvector associated with the positive eigenvalue, $v_1(z)$, obeys the $\mathbb{Z}_2$ symmetry, as does the eigenmode $v_2(z)$.
The other irrelevant eigenmode, however, is an even function, breaking the $\mathbb{Z}_2$ symmetry but nonetheless not affecting the leading order critical behavior of the network.
We omit the exact expressions of the eigenvalues and eigenvectors, as they are rather unwieldy and offer little insight, but for higher order truncations we plot the correlation length exponent versus dimension in Fig.~\ref{fig:IMbigfig}B and the eigenmodes in $d=3$  in Fig.~\ref{fig:IMbigfig}C.

For higher order truncations of the power series we find that the estimates of $\nu_\ast$ appear to converge as we increase the order of the truncation, but eventually become more and more difficult to solve numerically.
This said, our estimates close to $d = 3$, $\nu_\ast = 0.7$ at $\mathcal O(z^{25})$ match reasonably well with the known exponent of the Ising model universality class, $\nu_\ast = 0.63$.

\subsubsection{The effective nonlinearity near and at criticality}
\label{sec:criticalnonlin}

Finally, to connect back to the Widom scaling forms we clarify how the effective nonlinearity $\Phi(\psi)$ relates to the dimensionless nonlinearity $\varphi_1^\ast(z)$ and the eigenmodes $v_{1,\ell}(z)$. 
In doing so, we explain why the effective nonlinearity clearly depends on the bare nonlinearity $\phi(V)$, a non-universal property of the network, even when near or at a critical point. 

Consider a perturbation away from the fixed point, $\varphi_{1s}(z) = \varphi_1^\ast(z) + \delta \varphi_s(z)$. 
This perturbation is determined by the decomposition of the bare nonlinearity $\phi(V)$ into a series of eigenmodes near the fixed point:
\begin{align}
    \varphi_{1,s}(z) &= \varphi_1^\ast(z) + \sum_{\ell} c_\ell e^{\mu_\ell s} v_{1,\ell}(z),
    \label{eqn:eigenexpansion}
\end{align}
where $\mu_\ell$ are the eigenvalues, which are non-integer in general, $v_{1,\ell}(z)$ is the corresponding eigenmode to the $\varphi_{1,s}(z)$ component, and $c_\ell$ are the loadings onto each of these eigenmodes. 
In order for the model to be at a critical point the loadings $c_{\ell}$ corresponding to the relevant directions with positive eigenvalues $\mu_\ell > 0$ must be tuned to zero in order for the RG dynamics to lie on the critical manifold. 
The irrelevant directions, corresponding to negative eigenvalues $\mu_\ell < 0$, will decay as $s \rightarrow \infty$. 
At the critical points we focus on in this work there are two modes with positive eigenvalues. 
The first is the trivial constant mode $v_{1,0}(z) \propto 1$, which only contributes to the running baseline rate $\Pi_s$ and will not drive the dimensionless nonlinearity further away from the critical manifold.
The loading $c_1$ of the other mode with positive eigenvalue determines how close the network is tuned to the critical point, and the ``correlation length exponent'' $\nu_\ast$ is derived from this eigenvalue by the definition $\mu_1 \equiv (2\nu_\ast)^{-1}$.
The flow of the nonlinearity only depends on the synaptic weight $J$ (through the maximum eigenvalue $\Lambda_{\rm max}$), not $\mathcal E$, so we expect $c_1 \propto J_c-J$ close to the critical point. 

The effective nonlinearity $\Phi(\psi)$ is related to the dimensionless nonlinearity by
\begin{align*}
    &\Phi(\psi) - \nu_c - \Lambda_{\rm max}^{-1} (\psi - \psi_c) \\
    &= \lim_{s \rightarrow \infty} e^{-s(d/2+1-\eta^X_\ast)}\varphi_{1,s}\left((\psi-\psi_c)e^{s(d/2-\eta^X_\ast)}\right).
\end{align*}
In \emph{in vitro} networks both $\Pi_s \rightarrow \nu_c$ and $\theta_s \rightarrow \psi_c$ may be set to $0$.
To see how the RG critical point shapes the effective nonlinearity, we first plug in the expansion (\ref{eqn:eigenexpansion}).
\begin{align*}
    &\Phi(\psi) - \nu_c - \Lambda_{\rm max}^{-1} (\psi - \psi_c) \\
    &\sim e^{-s(d/2+1-\eta^X_\ast)}  \varphi_{1s}\left((\psi-\psi_c)e^{s(d/2-\eta^X_\ast)}\right)\\
    &= e^{-s(d/2+1-\eta^X_\ast)} \\
    & ~~~~ \times \Bigg\{c_1 e^{\frac{s}{2\nu_\ast}}v_1\left((\psi-\psi_c)e^{s(d/2-\eta^X_\ast)}\right) \\
    & \hspace{2.0cm} + \varphi_1^\ast\left((\psi-\psi_c)e^{s(d/2-\eta^X_\ast)}\right) \\
    & \hspace{2.0cm} + \sum_{\ell \geq 2} c_\ell e^{\mu_\ell s} v_{1,\ell}\left((\psi-\psi_c)e^{s(d/2-\eta^X_\ast)}\right)\Bigg\},
\end{align*}
where we have separated out the $\ell=1$ term from the eigenmodes.
We can understand the origin of scaling by imagining that for $c_1 \propto J_c - J \neq 0$ we only run the RG flow out to an RG-time $s$ such that $|J_c-J|e^{\frac{s}{2\nu_\ast}} = {\rm const.}$
We can thus replace $e^s \propto |J_c-J|^{-2\nu_\ast}$ in the eigenmode expansion above, which yields
\begin{widetext}
\begin{align}
    \Phi(\psi) - \nu_c - \Lambda_{\rm max}^{-1} (\psi - \psi_c) &\sim |J_c - J|^{2\nu_\ast(d/2+1-\eta^X_\ast)}f^\ast \left((\psi-\psi_c)|J_c - J|^{-2\nu_\ast(d/2-\eta^X_\ast)}\right) \label{eqn:effectivenonlinscaling} \\
    & ~~~~ + \sum_{\ell \geq 2} c_\ell |J_c - J|^{2\nu_\ast(d/2+1-\eta^X_\ast - \mu_\ell) } v_{1,\ell}\left((\psi-\psi_c)|J_c - J|^{-2\nu_\ast(d/2-\eta^X_\ast)}\right), \nonumber
\end{align}
\end{widetext}
where $f^\ast(z) \equiv {\rm const.} \times v_1(z) + \varphi_\ast(z)$.
To take stock of what we have derived, Eq.~(\ref{eqn:effectivenonlinscaling}) relates the effective nonlinearity $\Phi(\psi)$---which depends on the details of the bare nonlinearity $\phi(V)$---to the detail-independent universal properties of the RG fixed point: the fixed point function $\varphi_\ast(z)$, eigenmodes $v_\ell(z)$, and the critical exponents $\nu_\ast$, $\eta^X_\ast$, and $\mu_\ell$.
These universal quantities depend only on the properties of the RG critical point. 
The non-universal terms on the right-hand-side of Eq.~(\ref{eqn:effectivenonlinscaling}) are the loadings $c_\ell$, which describe the initial projection of the bare nonlinearity $\phi(V)$ on to the eigenmodes of the RG critical point \footnote{There is also information about the initial conditions contained in the constant prefactors of the running scales, which we have suppressed in Eq.~(\ref{eqn:effectivenonlinscaling}).}.
Thus, we see that non-universal information about the microscopic features of the model enter through the loadings $c_\ell$ associated with the corrections to scaling, and ultimately shape the effective nonlinearity $\Phi(\psi)$.
Importantly, although the RG critical point possesses an emergent $\mathbb{Z}_2$ symmetry, this symmetry is only inherited by $\varphi_1^\ast(z)$, not the eigenmodes $v_{1,\ell}(z)$, due to the fact that the bare action is not $\mathbb{Z}_2$ symmetric.
This is why the nonlinearities $\Phi(\psi) - \nu_c$ we observe in our simulations and by solving Eq.~(\ref{eqn:Phi1flow}) are not odd functions in $\psi-\psi_c$, even though the bare nonlinearity $\phi(V) - \phi(\theta)$ is an odd function in $V-\theta$.

As the network is tuned to its critical point at $J = J_c$, the critical nonlinearity retains its dependence on the non-universal $c_\ell$ and its lack of an overall $\mathbb{Z}_2$ symmetry.
One can show the asymptotic behaviors $\varphi_\ast(z) \sim \mathcal A_d z^{1 + \frac{1}{d/2-\eta^X_\ast}}$ and $v_{1,\ell}(z) \sim \mathcal B_\ell z^{1 + \frac{1-\mu_\ell}{d/2-\eta^X_\ast}}$ as $|z| \rightarrow \infty$, and the factors of $J_c - J$ ultimately cancel out of Eq.~(\ref{eqn:effectivenonlinscaling}) to yield the critical nonlinearity
\begin{align}
& \Phi(\psi) - \nu_c - \Lambda_{\rm max}^{-1} (\psi - \psi_c) \label{eqn:criticalnonlin}\\
&\sim \mathcal A_d (\psi-\psi_c)^{1 + \frac{1}{d/2-\eta^X_\ast}} + \sum_{\ell \geq 2} c_\ell \mathcal B_\ell (\psi-\psi_c)^{1 + \frac{1+|\mu_\ell|}{d/2-\eta^X_\ast}}, \nonumber
\end{align}
which still depends on the non-universal loadings $c_\ell$, but is now a series in \emph{non-analytic} powers of $\psi-\psi_c$.
For $\psi-\psi_c < 0$ these powers should be interpreted as the power of the absolute value of $|\psi-\psi_c|$, with an overall sign that depends on whether $v_{1,\ell}(z)$ is odd or even.

Plugging Eq.~(\ref{eqn:effectivenonlinscaling}) into the dynamical equation (\ref{eqn:truemeandyn}) for the mean membrane potential $\psi(t)$ and keeping only the leading scaling function $f^\ast(z)$ (the subleading terms contribute \emph{corrections to scaling} \cite{GoldenfeldBook1992,caillol2012non}), we obtain an implicit solution for $\psi(t) - \psi_c$ that can be formally inverted to give the Widom scaling forms (Eqs.~(\ref{eqn:widomscalingform})-(\ref{eqn:doublescalingform})) with $\beta_\ast = \frac{\nu_\ast d}{2}$ and $\Delta_\ast = \frac{\nu_\ast}{2}(d+4)$ in \emph{in vitro} networks and $\beta_\ast = \frac{\nu_\ast}{2}(d-2)$ and $\Delta_\ast = \frac{\nu_\ast}{2}(d+2)$ in \emph{in vivo} networks.
Comparing these expressions with the general scaling relations expected in the directed percolation ($\beta_\ast = \frac{\nu_\ast}{2}(d+\eta_\ast)$, $\Delta_\ast = \frac{\nu_\ast}{2}(d + 2 z_\ast - \eta_\ast)$) \cite{janssen2005field} and Ising ($\beta_\ast = \frac{\nu_\ast}{2}(d-2+\eta_\ast)$ and $\Delta_\ast = \frac{\nu_\ast}{2}(d+2 - \eta_\ast)$) universality classes, we confirm that our LPA approximation predicts trivial (mean-field) values of $z_\ast = 2$ and $\eta_\ast = 0$ for the dynamical and anomalous exponents \footnote{For the \emph{in vivo} networks, if we worked to higher order in $\tilde{z}$ we would obtain non-zero contributions to the anomalous scaling, giving $\beta_\ast = \frac{\nu_\ast}{2}(d-2-4\delta\eta^X_\ast)$ and $\Delta_\ast = \frac{\nu_\ast}{2}(d+2-4\delta\eta^X_\ast)$, where $\delta \eta^X_\ast = \eta^X_\ast - (d+2)/4$. The anomalous terms appear identically in the exponents, which means that in one of the exponent relations it has an incorrect sign. Since we find $\delta\eta^X_\ast > 0$, the sign is incorrect in $\beta_\ast$. The resolution to this issue is to go beyond the local potential approximation.}.

Because we identified the critical points we have found within the LPA with the well-known universality classes of directed percolation and the Ising model, in performing our scaling collapses in Sec.~\ref{sec:applications} we were able to use the known values of the critical exponents, including the non-trivial values of the anomalous exponent $\eta_\ast$ and dynamical exponent $z_\ast$, for the $2d$ and $3d$ lattices.
In our excitatory-inhibitory networks with excitatory random regular connections we used our LPA estimates as starting points for determining the values of the critical exponents that collapsed our data.
This demonstrates that even when networks lack translation invariance this may not drastically change the critical exponents.
That said, higher degree random regular networks with global inhibition proved to have mean-field scaling, suggesting that proper treatment of effects beyond the LPA are necessary to determine whether these networks can again be tuned to have anomalous scaling.

\end{singlespace}

%%%%%%%%%%%%%%%%%%%%%%%%%%%%%%%%
%% REFERENCES 
%%%%%%%%%%%%%%%%%%%%%%%%%%%%%%%%
%

%\bibliographystyle{unsrt}
%\bibliography{main}

\end{document}